\UseRawInputEncoding

\documentclass[aps,prl,amsmath,amssymb,floatfix,reprint,citeautoscript,noeprint,superscriptaddress,twocolumn]{revtex4-2}

\usepackage{dsfont}
\usepackage{lipsum} 
\usepackage{bibentry}
\usepackage[english]{babel}
\usepackage[dvipsnames]{xcolor}
\usepackage{graphicx}
\usepackage[caption=false]{subfig} 
\usepackage{amsmath,amssymb,bm}
\usepackage[version=3]{mhchem}
\usepackage{verbatim}
\usepackage{multirow}
\usepackage{dcolumn}
\usepackage{float}
\usepackage{nicefrac}
\usepackage{siunitx}
\usepackage{booktabs}
\usepackage{chemformula}
\usepackage{wrapfig}
\usepackage{enumitem}  
\usepackage{transparent}
\usepackage[colorlinks,allcolors=black,citecolor=blue,urlcolor=blue]{hyperref}
\usepackage{tcolorbox}
\usepackage{relsize}
\hypersetup{
    citecolor=Turquoise,
    colorlinks=true,
    linkcolor=black,    
    urlcolor=Turquoise,
    }

\newcommand{\mbf}[1]{\boldsymbol{\mathit{#1}}}
\newcommand{\mrm}[1]{\mathrm{#1}}
\newcommand{\mcl}[1]{\mathcal{#1}}
\newcommand{\tcr}[1]{\textcolor{black}{#1}}

\newcommand{\etal}{\emph{et al.}}

\newcommand{\erfc}{\mathrm{erfc}}
\newcommand{\erf}{\mathrm{erf}}

\usepackage{sansmathfonts}
\usepackage[justification=centerlast, font={sf}, labelfont={bf}]{caption}
\begin{document}

\title{Learning classical density functionals for ionic fluids}

\author{Anna T. Bui}
\affiliation{Yusuf Hamied Department of Chemistry, University of
  Cambridge, Lensfield Road, Cambridge, CB2 1EW, United Kingdom}

\author{Stephen J. Cox}
\email{stephen.j.cox@durham.ac.uk}
\affiliation{Department of Chemistry, Durham University, South Road,
  Durham, DH1 3LE, United Kingdom}


\date{\today}


\begin{abstract}
Accurate and efficient theoretical techniques for describing ionic
fluids are highly desirable for many applications across the physical,
biological and materials sciences. With a rigorous statistical
mechanical foundation, classical density functional theory (cDFT) is
an appealing approach, but the competition between strong Coulombic
interactions and steric repulsion limits the accuracy of current
approximate functionals. Here, we extend a recently presented machine
learning (ML) approach [Samm\"uller \etal,
Proc. Natl. Acad. Sci. USA, \textbf{120}, e2312484120 (2023)] designed
for systems with short-ranged interactions to ionic fluids. By
adopting ideas from local molecular field theory, the framework we
present amounts to using neural networks to learn the local
relationship between the one-body direct correlation functions and
inhomogeneous density profiles for a ``mimic'' short-ranged system,
with effects of long-ranged interactions accounted for in a
mean-field, yet well-controlled, manner. By comparing to results from
molecular simulations, \tcr{we show that our approach accurately
describes the structure and thermodynamics of prototypical models for
electrolyte solutions and ionic liquids, including size-asymmetric and
multivalent systems.} The framework we present acts as an important
step toward extending ML approaches for cDFT to systems with accurate
interatomic potentials.
\end{abstract}

\maketitle


The behavior of ionic fluids underlies a vast
array of physical and biological phenomena as well as
technological applications, ranging from
electrolyte solutions controlling protein folding 
\cite{Kunz2004} to room-temperature ionic liquids
in energy storage devices \cite{Kondrat2023}. 
A fundamental topic that continues to attract
enormous interest both experimentally
\cite{Espinosa-Marzal2014, Gebbie2015, Smith2016,
Lee2017, Gebbie2017} and theoretically
\cite{Torrie1980,  
Merlet2014, Bazant2011, Coupette2018, deSouza2020, Cats2021} is the structure and thermodynamics
of ions near charged interfaces, in particular, how the nature of both
the electrolyte and solid surface impacts the properties of the
electric double layer (EDL). Poisson--Boltzmann (PB) theory
and its linearized Debye-H\"{u}ckel form provide the 
basis for much of our understanding of ionic fluids. 
Their neglect of correlations arising
from non-electrostatic interactions, however, restricts their validity
to fluids of low ionic strength.


Classical density functional theory (cDFT) \cite{Evans1979, EvansBook,
Lutsko2010, HansenMcDonaldBook} provides a natural framework for
including correlations omitted by PB theory, and has proven to be a
powerful approach to describe equilibrium structure and
thermodynamics of fluids in general. While in principle an exact
theory, historically, cDFT relies on making approximations for the
excess intrinsic free energy functional
$\mcl{F}^{\mathrm{(ex)}}_{\mrm{intr}}[\{\rho_\nu\}]$, where $\rho_\nu$
denotes the one-body density of species $\nu$.  For
hard sphere fluids, functionals based on Rosenfeld's fundamental
measure theory (FMT)
\cite{Rosenfeld1989, Roth2002, Hansen-Goos2006, Roth2010}
have proven highly successful.  For simple liquids with square-well,
Yukawa, or Lennard--Jones interaction potentials, hard sphere mixtures
act as suitable reference systems, with effects of attractive interactions
described reasonably well in a mean-field fashion
\cite{Kierlik1991, Stewart2005, Archer2017}.
In contrast, existing functionals for
ionic fluids are far less accurate, failing to adequately capture the
interplay between Coulombic and steric interactions
\cite{Hartel2015, Hartel2017, Roth2016, Bultmann2022}. In this Letter,
we present a strategy that utilizes machine learning (ML) to construct
accurate free energy functionals for ionic fluids.

\begin{figure*}[t!]
  \includegraphics[width=\linewidth]{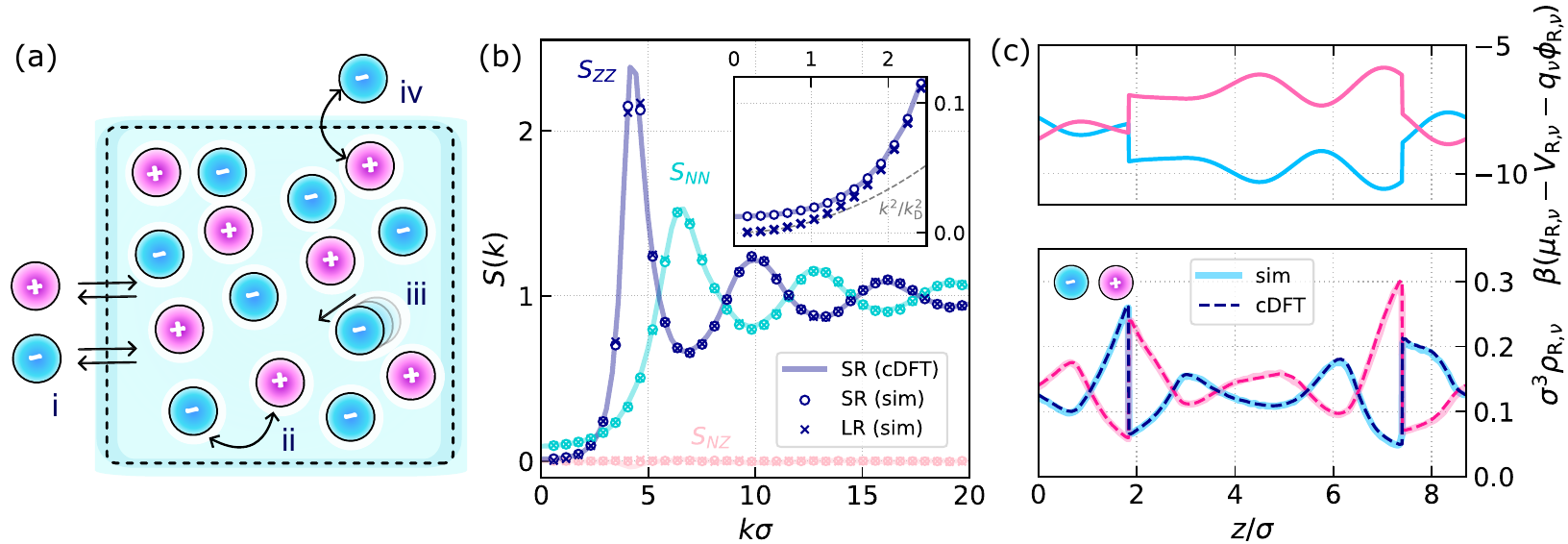} \caption{\textbf{Structure
  of the SR mimic RPM.}  (a) Training data for the neural functional
  are obtained from GCMC simulations with (i) insertion/deletion, (ii)
  position swapping, (iii) displacement and (iv) mutation (identity
  exchange) moves. (b) The static structure factors from cDFT agree
  well with results from molecular simulation across the whole range
  of $k$ for the SR system, shown for a bulk system with $\sigma^3\rho_{\mrm{R},\pm}=0.315$.
  Inset: At low $k$, the SR system violates
  the perfect screening condition, whereas the LR system obeys the
  Stillinger--Lovett sum rule (see Eq.~\ref{eqn:StillingerLovett}).
  (c) For the applied external potentials
  (top), predictions of the ion density profiles (bottom) are in
  excellent agreement with the simulation data.}
\label{fig1}
\end{figure*}

The rapid advance of modern ML approaches means there
has been much recent interest in ``learning'' 
representations of the \emph{exact}
$\mcl{F}^{\mathrm{(ex)}}_{\mrm{intr}}[\{\rho_\nu\}]$ \cite{Cats2021ml,
Dijkman2024, simon2024, Lin2020, Shang-Chun2019,
Malpica-Morales2023, Pederson2022}. Here we build on the method proposed by
Samm{\"u}ller \etal{} \cite{Sammuller2023}, in which the one-body direct
correlation functions,
 \begin{equation}
    c^{(1)}_{\nu}(\mbf{r};[\{\rho_\nu\}])= -\frac{\beta \delta \mcl{F}^{\mathrm{(ex)}}_{\mrm{intr}}[\{\rho_\nu\}] }{\delta \rho_\nu(\mbf{r})},
\end{equation}
are learned by generating inhomogeneous density profiles in the
presence of various external- and chemical potentials by grand
canonical (GC) simulations. This approach to cDFT, dubbed ``neural
functional theory,'' has been shown to outperform FMT-based approaches
for both hard sphere fluids and Lennard--Jones liquids in terms of
both accuracy and speed \cite{Sammuller2023,sammuller2024b}.

Two key features underpin the success of the neural functional
approach: \tcr{(i) correlations in the fluid are short-ranged (SR),
such that the functional relationship between $c_\nu^{(1)}$ and
$\{\rho_\nu\}$ is local; and (ii) the feasibility of GC simulations to
produce inhomogeneous density profiles of sufficient quality to form a
reliable training set.  Ionic fluids pose challenges on both
fronts. First, the long-ranged (LR) nature of the Coulomb potential
leads to a nonlocal relationship between $c_\nu^{(1)}$ and
$\{\rho_\nu\}$.  Second,} GC simulations for ionic systems raise
delicate questions concerning electroneutrality, with various schemes
proposed that either insert individual ions or neutral pairs
\cite{Shelley1994,Yan1999,Moucka2011,
Moucka2015, Kim2023}. Even then, simulations corresponding
exactly to the GC ensemble where the number of each species can
fluctuate are impractical, owing to poor acceptance rates of trial
insertion and deletion moves.  We circumvent both of these issues
by adopting concepts from local molecular field theory
(LMFT) \cite{Rodgers2008,Weeks1998, Rodgers2006, Chen2004,
Remsing2016}, which has been shown to have close links to
cDFT \cite{Archer2013}.

We initially focus our efforts on the prototypical model of an
ionic fluid: the restricted primitive model (RPM) comprising
oppositely charged hard spheres of equal diameter $\sigma$, embedded
in a uniform dielectric continuum. \tcr{Later, we will also present
results for a primitive model (i.e., size-asymmetric) and a
multivalent system.}  The scheme we propose can be briefly
summarized. First, by employing neural networks, we find the one-body
direct correlation functions for a suitably chosen ``mimic system''
whose electrostatic interactions are entirely short-ranged. Then, by
leveraging LMFT and its relation to cDFT, we account for the net
averaged effects of LR electrostatic interactions in a well-controlled
fashion. When compared to molecular simulations, the framework we
outline describes inhomogeneous density profiles, the equation of
state, and the properties of the electric double layer in the presense
of electric fields with very high accuracy.

\textit{\textbf{cDFT of a short-ranged ``mimic'' ionic fluid.}}
Our overall strategy follows that of LMFT, in which we consider a
suitably chosen mimic system whose interatomic interactions are
entirely short-ranged, and, when subject to a suitably chosen one-body
potential $\phi_{\rm R}$, has the same one-body densities as the system
of interest with LR interactions (the ``full'' system). For systems
such as the RPM, where LR interactions arise from the Coulomb
potential, one adopts the exact splitting
\begin{equation}
\label{eqn:splitting}
    1/r = v_0(r) + v_1(r),
\end{equation}
with $v_0(r) = \erfc(\kappa r)/r$ and $v_1(r) = \erf(\kappa r)/r$. The
interatomic potential of the mimic system is then $v_0$, and the
length scale $\kappa^{-1}$ is chosen such that the mimic system
accurately describes the SR correlations of the full system. In the
following, we will work with $\kappa^{-1}=1.8\,\sigma$ (see the SM \cite{supp}).  We
will also indicate quantities pertaining to the mimic system with a
subscript ``R'' and focus on behaviors at a reduced temperature
$T^*=0.066$, corresponding to supercritical conditions
\footnote{The critical temperature of the RPM is 
$T^*_{\mrm{c}}=0.0492$ (from Ref.~\onlinecite{Yan1999})}.  The
one-body direct correlation functions of the mimic system can be
exactly written as
\begin{equation}
    c^{(1)}_{\mrm{R},\nu}(\mbf{r}) = \ln \Lambda_{\nu}^3\rho_{\mrm{R},\nu}(\mbf{r}) + \beta V_{\mrm{R},\nu}(\mbf{r}) + \beta q_\nu \phi_{\mrm{R}}(\mbf{r}) - \beta \mu_{\mrm{R},\nu},
    \label{eqn:c1_mimic}
\end{equation}
where $\Lambda_{\nu}$, $\mu_{\mrm{R},\nu}$ and $q_\nu$ indicate the
thermal de Broglie wavelength, chemical potential and point
charge of species $\nu$, respectively, $V_{\mrm{R},\nu}$ encompasses
any non-electrostatic contributions to the external potential for
species $\nu$ and, for now, $\phi_{\mrm{R}}$ is a general external
electrostatic potential; we will later discuss how $\phi_{\rm R}$ can
be chosen such that $\rho_{\mrm{R}, \nu}(\mbf{r})
= \rho_{\nu}(\mbf{r})$.

To learn the functional relationship for
$c^{(1)}_{\mrm{R},\nu}(\mbf{r};[\{\rho_{\mrm{R},\nu}\}])$, we obtain
data for the right hand side of Eq.~\ref{eqn:c1_mimic} by measuring
$\rho_{\mrm{R},\nu}$ from simulations with known $\beta
V_{\mrm{R},\nu}$, $\beta \phi_{\mrm{R}}$ and
$\beta \mu_{\mrm{R},\nu}$.  To this end, in line with
Ref.~\onlinecite{Sammuller2023}, we perform \tcr{GC} simulations in a planar
geometry at different combinations of $\{\beta V_{\mrm{R},\nu}\}$,
$\beta \phi_R$ and $\{\beta \mu_{\mrm{R},\nu}\}$. \tcr{Note that GC simulations are essential for 
the purpose of evaluating Eq.~\ref{eqn:c1_mimic} and, on their own, canonical methods 
such as molecular dynamics (MD) simulations are insufficient.} 
The form of $v_0$ means that each particle of the mimic system can be considered
electroneutral, comprising both a point charge and a compensating
Gaussian charge distribution \cite{Frenkel2023}.  As such, in addition
to translational moves, we can readily perform GC particle
insertions/deletions, along with semi-GC swapping of ``anions'' and
``cations'' without needing to worry about issues of electroneutrality
[see Fig.~\ref{fig1}(a)]. We have found this highly beneficial to
converging our simulations, full details of which are provided in the
SM \cite{supp}. \tcr{In total, $\sim 2500$  simulations have been performed to gather training data.}

For each species, $\nu=+\,\mrm{or}\,-$, the neural network used to
represent the relationship $\{\rho_{\mrm{R},\nu}(z) \} \rightarrow
c^{(1)}_{\mrm{R},\nu}(z)$ is structured as follows. The input layer
has two channels \tcr{that} are supplied with the discretized values
of the cation and anion density profiles in a window $\Delta
z=3.6\,\sigma$ around the particular value of $z$.  Following a
fully-connected multilayer perceptron of three layers, the output
layer consists of a single node, which yields the predicted value of
$c^{(1)}_{\mrm{R},\nu}$ at position $z$. For each ionic fluid considered, we
train two independent networks, one for the cation and one for the
anion. To reduce noise in the bulk structure predictions, we also
obtained models regularized with bulk two-body direct correlation
functions as proposed in Ref.~\onlinecite{sammuller2024}. Details of
our training procedure are provided in the SM \cite{supp}, and all
data are openly available \cite{zenodo}.

\begin{figure*}[t!]
  \includegraphics[width=\linewidth]{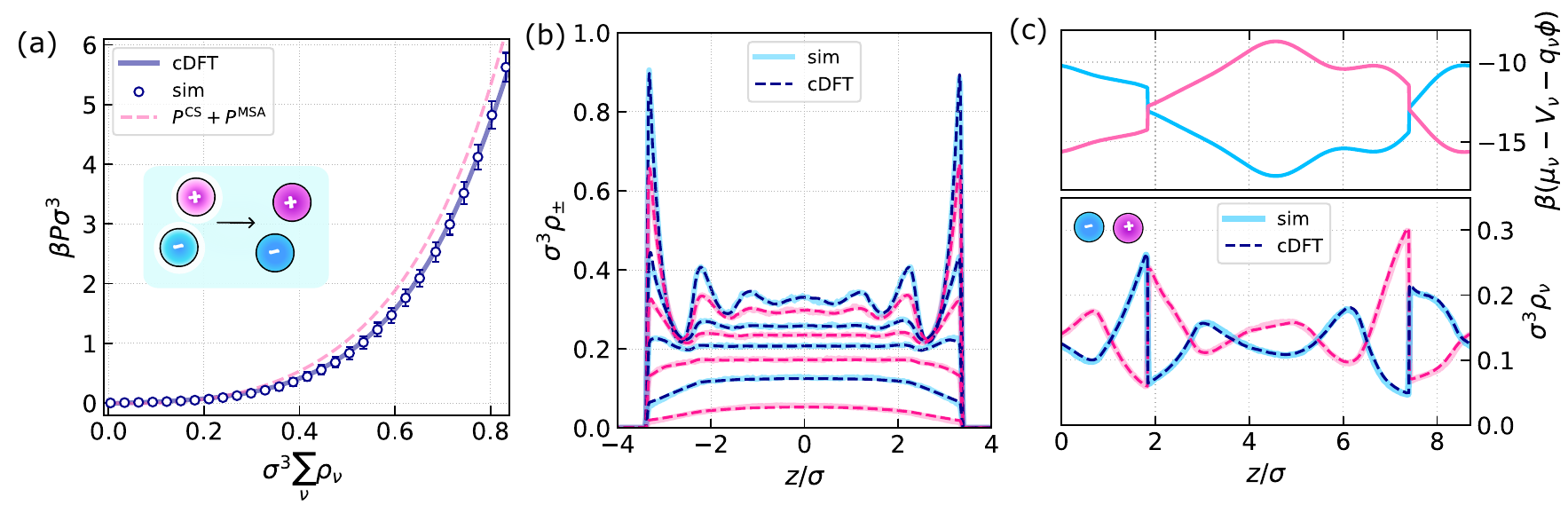} \caption{\textbf{Structure
  and thermodynamics of the LR full system.} cDFT for the RPM at
  $T^*=0.066$ with effects of LR electrostatic interactions accounted
  by LMFT shows excellent agreement with reference molecular
  simulation data for (a) the equation of state, (b) ion density
  profiles confined in a slit and (c) with an applied external
  potential. In (a), we see that our cDFT approach outperforms the
  prediction obtained by combining the Carnahan--Starling equation of
  state for hard-sphere fluids ($P^{\mrm{CS}}$) and the mean spherical
  approximation ($P^{\mrm{MSA}}$).  For the LR system in (c), the
  applied external potential acts over a much longer range and the
  chemical potentials are shifted lower compared to the SR mimic
  system in Fig.~1(c) to yield $\rho_\nu(z) = \rho_{\mrm{R},\nu}(z)$.}
\label{fig2}
\end{figure*}

We first assess the bulk structure predicted by the functional.  On
the basis of the neural network, we make use of automatic
differentiation and radial projection to obtain the partial two-body
direct correlation functions $c^{(2)}_{\nu\lambda}(r)$ (see
Ref.~\cite{Sammuller2023} and the SM \cite{supp}).  By solving the general
Ornstein--Zernike equation for mixtures
\cite{Ornstein1914, Baxter2003},
we then obtain the total correlation functions
$h^{(2)}_{\nu\lambda}(r)$ and the corresponding structure factors
$S_{\nu\lambda}(k)$. We further quantify the degree of coupling
between number--number ($S_{NN}$), number--charge ($S_{NZ}$) and
charge--charge ($S_{ZZ}$) densities by appropriate weighted
summations
\begin{equation}
\begin{split}
    S_{NN}(k) &= S_{++}(k) + 2S_{+-}(k) + S_{--}(k), \\
    S_{NZ}(k) &= S_{++}(k)  - S_{--}(k), \\
    S_{ZZ}(k) &= S_{++}(k) - 2S_{+-}(k) + S_{--}(k).
\end{split}
\label{eqn:Sk}
\end{equation}

Fig.~1(b) shows that the static structure factors obtained from the
functional are in excellent agreement with results from a molecular
dynamics simulation of the mimic system at the same bulk
densities \footnote{\tcr{While GC simulations (i.e., known $\{\mu_\nu\}$) are required 
to train the neural networks (see Eq.~\ref{eqn:c1_mimic}), it is reasonable to 
compare to average structures obtained with canonical simulations at the same average density.}}.  
The symmetry of the RPM is well-captured, reflected in $S_{NZ}(k)=0$.
Moreover, by comparing to results from a MD simulation of the full
system, we see that $S_{NN}(k)$ and $S_{ZZ}(k)$ for the full and mimic
systems agree very well at sufficiently large $k$, confirming that our
choice of $\kappa^{-1} = 1.8\,\sigma$ is sufficient for the mimic
system to capture the SR correlations of the full system.  Deviations
of the SR system from the LR system only manifest significantly in
$S_{ZZ}(k)$ at small $k$, in agreement with previous works on the
subject \cite{Chen2004, Cox2020}. In particular, as shown in the
inset, we see that $S_{ZZ}(k)$ for the LR system strictly obeys the
Stillinger--Lovett sum rule
\cite{Stillinger1968a, Stillinger1968b}
\begin{equation}
   \lim_{k\rightarrow 0}\frac{k^2_{\rm D}}{k^2} S_{ZZ}(k) = 1,
   \label{eqn:StillingerLovett}
\end{equation}
where $k_{\rm D}^{-1}$ is the Debye screening length.
In contrast, for the mimic system we observe $S_{ZZ}(k) > k^2/k^2_{\rm
D}$ as $k\to 0$, indicative of a lack of screening.
Consistent with the dielectric response of a
short-ranged water model \cite{Cox2020}, these deviations appear at
length scales far larger than the range separation prescribed by
$2\pi\kappa=3.5\,\sigma^{-1}$.

Turning to the ability of cDFT to predict inhomogeneous structure, the
equilibrium density profiles of the mimic system can be obtained by
self-consistently solving the Euler--Lagrange equation
\begin{equation}
\begin{split}
    \Lambda^3_{\nu}\rho_{\mrm{R},\nu}(\mbf{r}) = \exp\Big(-\beta V_{\mrm{R},\nu}(\mbf{r}) & - \beta q_\nu \phi_R(\mbf{r})  \\
    & + \beta \mu_{\mrm{R},\nu} + c^{(1)}_{\mrm{R}, \nu}(\mbf{r}; [\{ \rho_{\mrm{R},\nu}\}]) \Big),
\end{split}
\end{equation}
where $c_{\mrm{R},\nu}^{(1)}$ is evaluated by the corresponding
neural network.  As
shown in Fig.~1(c) for a representative set of external potentials,
the resulting $\{\rho_{\mrm{R},\nu}(z)\}$ from cDFT are in excellent
agreement with reference simulation data of the mimic system.

\textit{\textbf{Accounting for long-ranged electrostatics with LMFT.}}
Having established that the cDFT obtained from the ML procedure is
accurate for a SR variant of the RPM, we turn our attention to
incorporating the effects of LR electrostatic interactions. To do so, we
will use concepts from LMFT.
The premise of LMFT is that there exists a
potential $\phi_{\rm R}$ such that
\begin{equation}
 \rho_\nu (\mbf{r}; [\phi], \{\mu_{\nu}\}) =  \rho_{\mrm{R},\nu} (\mbf{r}; [\phi_{\mrm{R}}], \{\mu_{\mrm{R},\nu}\} ).
 \label{eqn:density-equal}
\end{equation}
In Eq.~\ref{eqn:density-equal}, we have explicitly indicated the
functional dependence of the densities on the external electrostatic potential
\footnote{The number densities also
have a functional dependence on the non-electrostatic contribution to
the external potential. As this is the same for both the full and
mimic systems, i.e $\{V_{\nu}\}=\{V_{\nu,\mrm{R}}\}$, we do not indicate this functional dependence explicitly in Eq.~\ref{eqn:density-equal}.}. 
As shown in Ref.~\onlinecite{Archer2013}, 
when recast in a cDFT framework, LMFT
relates the one-body direct correlation 
functions of the full and mimic systems through
\begin{equation}
    c_{\nu}^{(1)}(\mbf{r}; [\{ \rho_{\nu}\}]) =
    c_{\mrm{R},\nu}^{(1)}(\mbf{r}; [\{ \rho_{\nu}\}])
    - \beta \Delta \mu_{\nu} + \beta q_\nu \Delta \phi(\mbf{r}),
    \label{eqn:cDFT-LMFT}
\end{equation}
which is an exact result. A key insight from LMFT
\cite{Weeks1998, Rodgers2008}
is that for an appropriate splitting of the potential (see
Eq.~\ref{eqn:splitting}), $\Delta\phi$ is well-approximated by a mean
field form
\begin{equation}
   \Delta \phi  (\mbf{r}) \equiv \phi(\mbf{r}) - \phi_{\mrm{R}}(\mbf{r}) = - \frac{1}{\epsilon} \int\!\!\mrm{d}\mbf{r}^\prime\, n(\mbf{r}^\prime)\, v_1(|\mbf{r}-\mbf{r}^\prime|),
   \label{eqn:phi-correction}
\end{equation}
where $\epsilon$ is the dielectric constant of the continuum and
$n=n_{\mrm{R}}=\sum_{\nu}q_{\nu}\rho_{\nu}$ is the
charge density.
For homogeneous systems, Rodgers and Weeks 
have also derived a thermodynamic correction for the total energy,
from which a correction for the pressure follows \cite{Rodgers2009},
\begin{equation}
    \Delta P  \equiv P - P_R = - \frac{k_{B}T}{2 \pi^{3/2}\kappa^{-3}}.
\label{eqn:pressure-correction}
\end{equation}
Based on Ref.~\onlinecite{Rodgers2009}, we have further derived a
correction for the chemical potential (see the SM \cite{supp}),
\begin{equation}
  \Delta  \mu_{\nu} \equiv \mu_{\nu} - \mu_{\mrm{R},\nu} = -\frac{q^2_\nu}{\kappa^{-1}\sqrt{\pi}}.
  \label{eqn:mu-correction}
\end{equation}

{We first consider the bulk thermodynamics of the RPM.
The equation of state for the LR system is} obtained 
with 
\begin{equation}
\label{eqn:cDFT-EoS}
    P(\{\rho_{\nu}\})= \sum_{\nu }k_{\mrm{B}}T\rho_{\nu} \left(1- c^{(1)}_{\mrm{R},\nu}[\{\rho_{\nu}\}]\right)- \frac{\mcl{F}^{\mrm{(ex)}}_{\mrm{intr}, \mrm{R}}[\{\rho_{\nu}\}]}{V} + \Delta P,
\end{equation}
where the sum of the first two terms is the negative of the grand
potential density of the mimic system.  The excess free energy
$\mcl{F}^{\mrm{(ex)}}_{\mrm{intr}, \mrm{R}}$ is accessible by
functional line integration.  The result of this procedure, shown in
Fig.~\ref{fig2}(a), agrees very well with reference simulation
data. We also see that, particularly at higher densities,
Eq.~\ref{eqn:cDFT-EoS} performs significantly better than the
analytical approximation that results from adding the
Carnahan--Starling equation of state
($P^{\mrm{CS}}$) \cite{Carnahan1969} and the mean spherical
approximation ($P^{\mrm{MSA}}$) \cite{Lebowitz1966, Blum1975}, despite
the fact that $P^{\mrm{CS}}+P^{\mrm{MSA}}$ forms the basis for the
vast majority of current state-of-the-art functionals for ionic
fluids \cite{Roth2016, Bultmann2022}.

The advantages of cDFT combined with LMFT become even clearer when
inhomogeneous systems are considered.  The equilibrium
densities of the full system are given by
\begin{equation}
\begin{split}
   \Lambda^3_{\nu} \rho_{\nu}(\mbf{r}) = \exp\Big(-\beta V_{\nu}(\mbf{r}) & - \beta q_\nu  \phi(\mbf{r}) \\
    & + \beta \mu_{\nu}  + c^{(1)}_{\nu}(\mbf{r}; [\{ \rho_{\nu}\}]) \Big),
\end{split}
\label{eqn:EL-LR}
\end{equation}
where the non-local functional $c_{\nu}^{(1)}$ is now given by
Eq.~\ref{eqn:cDFT-LMFT}, together with Eq.~\ref{eqn:phi-correction}
and Eq.~\ref{eqn:mu-correction}. From a cDFT perspective, we have
captured all strong rapidly-varying short-ranged correlations
with the neural networks for $c_{\mrm{R},\nu}^{(1)}$, while the
effects of LR electrostatic interactions are incorporated in a
mean-field, yet well-controlled fashion.


In practical terms, for a planar geometry, Eq.~\ref{eqn:phi-correction}
can be recast as
\begin{equation}
    \Delta \phi(z) = - \frac{1}{\epsilon L}\sum_{k\neq 0} \frac{4\pi}{k^2} \tilde{n}(k)\exp(ikz) \exp\left(-\frac{k^2}{4 \kappa^2}\right),
\end{equation}
where $\tilde{n}$ denotes a Fourier component of $n$, and $L$ is the
total length of the periodic cell.  In Fig.~2(b), we show the
results from our cDFT approach for the RPM confined between two
repulsive walls for various values of $\mu_+ = \mu_-$.
%
%
The density profiles are in very good agreement with reference
canonical MD simulation data in which the number of particles matches
that obtained by integrating the density profiles from cDFT. Our cDFT
approach also works very well in cases where the external potential
contains an electrostatic component, as illustrated in
Fig.~2(c). Here, the full system is the LR counterpart of the mimic
system shown in Fig.~1(c), emphasizing the premise of LMFT (see
Eq.~\ref{eqn:density-equal}).

\textit{\textbf{The electric double layer and other ionic fluids.}}
The results presented so far demonstrate that our cDFT approach
provides an efficient and accurate route to the structure and
thermodynamics of the RPM. As an application, we turn our
attention to the fundamental topic of ongoing scientific interest: the
electric double layer. We also show that our approach is robust
to choice of interatomic potential. Specifically, we also consider the
 primitive model (PM) and a multivalent fluid.

As a simple model of an EDL-forming system, we consider the RPM
confined between two repulsive walls, and in the presence of an
electric field $E_z$ along the $z$ direction.
The electrostatic potential that the LR system feels is $\phi(z)=-E_z
z$.
For the reference simulation data, the LR system was simulated at
constant $E_z$ using the finite-field approach
\cite{Stengel2009, Zhang2016, Zhang2016b, Sprik2018, Cox2019, Sayer2020, Zhang2020}.
As shown in Fig.~3(a), the ion distributions from our cDFT approach
are in excellent agreement with the results from
simulation. Notably, the cDFT accurately describes
the strong oscillations in the density profiles of co- and
counter-ions. Moreover, as we detail in the SM \cite{supp}, our
functional is thermodynamically consistent, satisfying the contact
density theorem \cite{Henderson1978, Henderson1979, Martin1988}
\begin{equation}
\label{eqn:contact}
    P = -\sum_{\nu} \int\!\!\mrm{d}z\,\rho_{\nu}(z)\frac{\mrm{d}V_{\nu}(z)}{\mrm{d}z} -  \frac{2\pi \sigma_{\mrm{s}}^2 }{\epsilon},
\end{equation}
which relates the bulk pressure to the surface charge density
$\sigma_{\rm s} = \int_{-L}^0\!\mrm{d}z\,n(z)$ \footnote{This equation
for $\sigma_{\rm s}$ assumes that the electrolyte is centered in a
periodically replicated cell of length $L$ along the $z$ direction.}
and ion densities at contact with the
walls. Obeying this contact theorem has proven a challenge \tcr{not
only for integral equation methods
\cite{Ichiye1990, Nygard2013,Jing2015, Dinpajooh2024},
which tend to violate key
sum rules and are therefore generally
thermodynamically inconsistent 
\cite{HendersonBook, Archer2013},
but also} for most state-of-the-art
cDFT functionals for ionic fluids \cite{Voukadinova2018,
Gillespie2014}. \tcr{The accurate description of EDL structure and
thermodynamic consistency displayed by our LMFT-based neural cDFT
approach constitutes a significant advancement in the theoretical
description of ionic fluids.}


\begin{figure}[t]
  \includegraphics[width=0.9\linewidth]{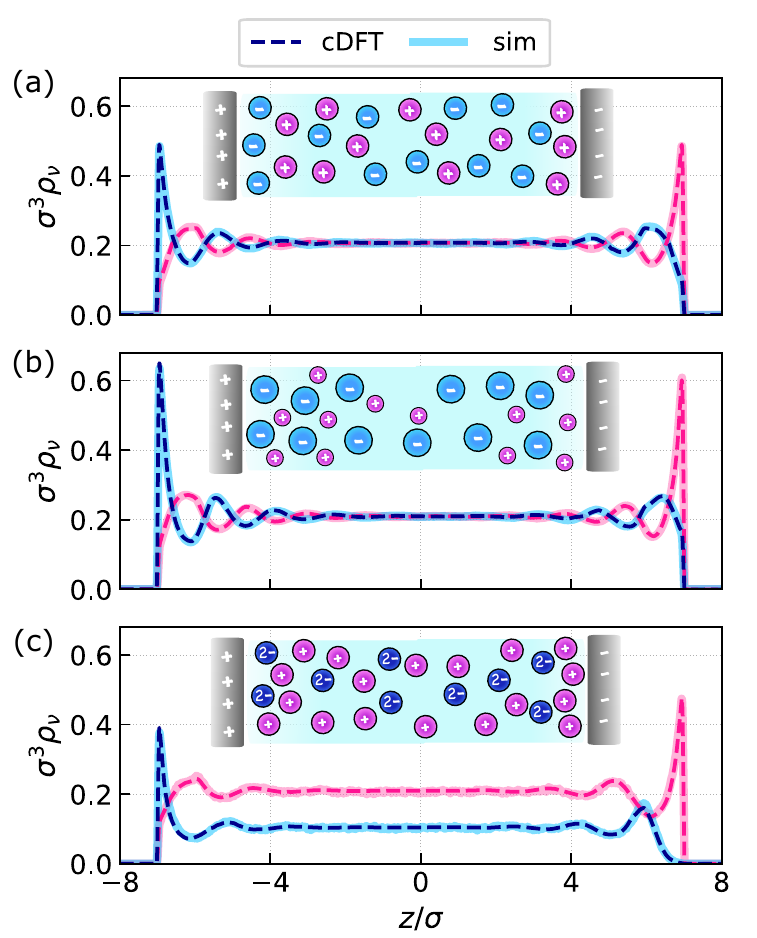} \caption{ \tcr{\textbf{Accurately
  describing the EDL with cDFT for (a) the RPM, (b) the PM, and (c) a
  multivalent fluid.}  In all cases, confining walls located at
  $\tcr{z/\sigma = \pm 7.2}$ in a periodic cell of $\tcr{L/\sigma=18.1}$, and
  $\tcr{E_z = 1.7 \,k_{\mrm{B}} T /e\sigma}$.} The resulting ion density profiles from cDFT are
  in excellent agreement with molecular simulations, and obey the
  contact theorem (Eq.~\ref{eqn:contact}).}  \label{fig3}
\end{figure}

The main advantage of the neural functional approach outlined in
Ref.~\onlinecite{Sammuller2023} is arguably that the local
relationship between $c_\nu^{(1)}$ and $\{\rho_{\nu}\}$ permits
application of the resulting functional to system sizes far beyond
those encountered during training of the ML. By using the framework of
LMFT and its relationship to cDFT, we have successfully extended the
neural functional approach to a case where the relationship between
$c_\nu^{(1)}$ and $\{\rho_{\nu}\}$ is nonlocal. To demonstrate
that this methodology is not limited to the RPM, in Figs.~3(b)
and~(c), we present results for a PM and multivalent system. For the PM, we have changed the sizes of the anion and
cation to $\sigma_- = 4\sigma/3$ and $\sigma_+ = 2\sigma/3$,
respectively. For the multivalent system, we consider equal-sized
anions and cations of diameter $\sigma$, but increase the valency of
the anion such that $q_-/q_+ = 2$. We use the same temperature as for
the RPM, $T^{*}= \epsilon\sigma k_{\mrm{B}}T/|q_{+}q_{-}|=0.066$. In
both cases, we observe excellent agreement between the neural cDFT and
simulations, both for inhomogeneous profiles and
bulk equations of state, as further shown in the SM \cite{supp}).  Future work will pursue using neural functionals to
understand EDL capacitance \cite{Cats2021b} and surface force balance
measurements \cite{Cats2021}, and to augment other cDFT approaches to
understand solvation \cite{Bui2024}. The framework we have outlined
should also find use in describing the dielectric response of polar
fluids such as water \cite{Gao2023, Rodgers2008b, Cox2020}.

\nocite{Panagiotopoulos1992, Thompson2022, Schneider1978,Archer2013, Hartel2016, 
Jover2012,
Brukhno2021, deLeeuw1997, hockney1988, Kolafa1992, Chollet2017,
sayer2017, sayer2019, Sedlmeier2011, Johnson1993, github2, github3}

\textit{\textbf{Data availability.}} Training data and code supporting the findings of this study will be openly available \cite{github, zenodo}.

\textit{\textbf{Acknowledgements.}} 
Via membership of the UK's HEC Materials
Chemistry Consortium funded by EPSRC
(EP/X035859), this work used the ARCHER2 UK
National Supercomputing Service. A.T.B. acknowledges funding from the Oppenheimer Fund and Peterhouse
College, University of Cambridge.  S.J.C. is a Royal Society
University Research Fellow (Grant No. URF\textbackslash
R1\textbackslash 211144) at Durham University.

\bibliography{references}

\end{document}


\title{Supplemental material: Learning classical density functionals for ionic fluids}

\author{Anna T. Bui}
\affiliation{Yusuf Hamied Department of Chemistry, University of
  Cambridge, Lensfield Road, Cambridge, CB2 1EW, United Kingdom}

\author{Stephen J. Cox}
\email{stephen.j.cox@durham.ac.uk}
\affiliation{Department of Chemistry, Durham University, South Road,
  Durham, DH1 3LE, United Kingdom}

\date{\today}

\maketitle

\tableofcontents
\newpage

\section{Classical density functional theory} 

For an $n$-component ionic fluid described by the singlet
ion densities $\{\rho_\nu(\mbf{r})\}$ ($\nu=1,2,...,n$) with
the charge carried by ions of species $\nu$ being $q_\nu$,
the grand potential functional is
\begin{equation}
    \varOmega[\{\rho_{\nu}\}] = \mcl{F}_{\mrm{intr}}[\{\rho_\nu\}] - \sum_\nu\int\!\mrm{d}\mbf{r}\,\left[\mu_\nu - V_{\nu}(\mbf{r}) - q_\nu \phi(\mbf{r})\right]\rho_{\nu}(\mbf{r}),
\end{equation}
where $\mu_\nu$ is the chemical potential of species $\nu$, $V_{\nu}$ contains all non-coulombic contributions to the external potential on species 
$\nu$, and $\phi$ is the electrostatic potential.
The intrinsic Helmholtz free energy functional $\mcl{F}_{\mrm{intr}}$  can be divided exactly into ideal and
excess parts 
\begin{equation}
   \beta \mcl{F}_{\mrm{intr}}[\{\rho_\nu\}] = \sum_{\nu} \!\int\!\mrm{d}\mbf{r}\,\rho_{\nu}(\mbf{r})\left[\ln\left(\Lambda_{\nu}^3\rho_{\nu}(\mbf{r})\right) -1\right]  + \beta\mcl{F}_{\mrm{intr}}^{\mrm{(ex)}}[\{\rho_\nu\}],
\end{equation}
with $\beta=(k_{\mrm{B}} T)^{-1}$, where $k_{\mrm{B}}$ is Boltzmann's constant, $T$ is the temperature and  
$\Lambda_{\nu}$ is the de Broglie thermal wavelength of species $\nu$.
The intrinsic excess free energy functional $\mcl{F}_{\mrm{intr}}^{\mrm{(ex)}}$ acts as a generating functional for the one-body direct correlation function
of each species $\nu$: 
 \begin{equation}
    c^{(1)}_{\nu}(\mbf{r};[\{\rho_\nu\}])= -\frac{\beta \delta \mcl{F}^{\mathrm{(ex)}}_{\mrm{intr}}[\{\rho_\nu\}] }{\delta \rho_\nu(\mbf{r})}.
\end{equation}
The variational principle states that the equilibrium ion
densities minimise $\varOmega$, so the 
Euler--Lagrange equation for each species $\nu$ is given as
\begin{equation}
   \rho_{\nu}(\mbf{r}) =  \Lambda_{\nu}^{-3}\exp\left(-\beta V_{\nu}(\mbf{r}) - \beta q_\nu  \phi(\mbf{r}) + \beta \mu_{\nu} + c^{(1)}_{\nu}(\mbf{r}; [\{ \rho_{\nu}\}]) \right).
\end{equation}
Upon rearrangement, we can write the relationship for learning
the one-body correlation functions
\begin{equation}
    c^{(1)}_{\nu}(\mbf{r}) = \ln\left( \Lambda_{\nu}^3 \rho_{\nu}(\mbf{r})\right) + \beta V_{\nu}(\mbf{r}) + \beta q_\nu  \phi(\mbf{r}) - \beta \mu_{\nu}.
\end{equation}

\section{Simulations of the short-ranged system}

The full system we considered is the restricted primitive model (RPM) in which the solvent is treated as a dielectric continuum with dielectric constant $\epsilon$, and the ions are modeled as hard spheres
of two species (cations and anions) that are characterized by
equal diameter $\sigma$ and opposite charges $q_{+}=-q_{-}$.
The pair potentials of the RPM are defined by
\begin{equation}
 u_{ij}(r) =
    \begin{cases}
      \infty & \quad   r\leq \sigma\\
      \dfrac{q_i q_j}{\epsilon } \dfrac{1}{r}  & \quad   r>\sigma.
    \end{cases} 
    \label{eqn:pair_potential_LR}
\end{equation}
Throughout, we adopt the Gaussian unit system for electrostatics in which $4\pi\epsilon_0= 1$ where $\epsilon_0$ is the permittivity of free space. The short-ranged mimic system has pair potentials
\begin{equation}
 u_{ij, \mrm{R}}(r) =
    \begin{cases}
      \infty & \quad   r\leq \sigma\\
      \dfrac{q_i q_j}{\epsilon } \dfrac{\mrm{erfc(\kappa r)}}{r} & \quad   \sigma < r \leq r_{\mrm{c}} \\
      0 & \quad   r>r_{\mrm{c}},
    \end{cases}       
    \label{eqn:pair_potential_SR}
\end{equation}
where $\kappa^{-1}=1.81\,\sigma$ and $r_{\mrm{c}}=3.6\,\sigma$.
It is also convenient to introduce a reduced unit system, in which the unit of energy is 
$|q_{\pm}|^2/\epsilon\sigma$.
Therefore, the reduced temperature is given as 
\begin{equation}
T^* = \dfrac{\epsilon  \sigma k_{\mathrm{B}} T}{|q_{+}q_{-}|}.
\end{equation}
In the article, we focus on the RPM at $T^*=0.066$ under supercritical condition.
In the main paper, we report quantities in their reduced units.
Here in the SM, to stay in line with how
the simulations and calculations
are performed in practice, we will report quantities in ``real'' units and 
we employed $\sigma=2.76\,\mrm{\AA}$, $q_{-}=-q_{+}=e$ where $e$ is the elementary charge and $\epsilon=1$, which corresponds to molten NaCl \cite{Panagiotopoulos1992} 
at $T=4000\,\mrm{K}$.

\subsection{GCMC simulations}

For the short-ranged systems, we performed GCMC simulations
with our own code \cite{github2}. Insertion and deletion moves were performed separately for each ionic species, i.e., $\mu_{\mrm{R},+}$ and $\mu_{\mrm{R}, -}$ were controlled independently.
Cubic simulation boxes of length $L=24\,\mrm{\AA}$ with periodic
boundary conditions were used. Each system was equilibrated for 
at least $1\times 10^5$ MC steps. In each MC move, a choice of the species $\nu$ is made randomly, followed by five possibilities with the following transition probabilities
from state $i$ to state $j$, $p_j/p_i$:
\begin{enumerate}[label=(\alph*)]
    \item Displacement of an ion of species $\nu$
    \begin{equation}
        \frac{p_j}{p_i} = \mrm{exp}(-\beta\Delta U_{ji}),
    \end{equation}
    \item Insertion of an ion of species $\nu$
    \begin{equation}
        \frac{p_j}{p_i} = \frac{V}{(N_{\nu}+1)\Lambda^{3}_{\nu}}\mrm{exp}(-\beta\Delta U_{ji} + \beta\mu_{\mrm{R},\nu}),
    \end{equation}
    \item Deletion of an ion of species $\nu$
    \begin{equation}
        \frac{p_j}{p_i} = \frac{N_{\nu} \Lambda^{3}_{\nu}}{V}\mrm{exp}(-\beta\Delta U_{ji} - \beta\mu_{\mrm{R},\nu}),
    \end{equation}
    \item Mutation of an ion of species $\nu$  to an ion of species  $\lambda$
    \begin{equation}
        \frac{p_j}{p_i} = \frac{N_{\nu} \Lambda^{3}_{\nu}}{(N_{\lambda}+1)\Lambda^{3}_{\lambda}}\mrm{exp}(-\beta\Delta U_{ji} - \beta\mu_{\mrm{R},\nu} +  \beta\mu_{\mrm{R},\lambda}),
    \end{equation}
    \item Swap position of an ion of species $\nu$ with an ion of species $\lambda$ 
    \begin{equation}
        \frac{p_j}{p_i} = \mrm{exp}(-\beta\Delta U_{ji}),
    \end{equation}
    where $N_{\nu}$ denotes the number of ions of species $\nu$,
    $V$ is the volume of the simulation box and
    $\Delta U_{ji}$ is the change in potential energy between state $i$ and state $j$, including any contribution from external potentials. 
    In all cases, a trial move was accepted with probability $\min(1,p_j g_{j\rightarrow i}/p_i g_{i\rightarrow j})$, where $g$ is the generation probability by drawing a random number uniformly in the interval $[0,1)$.
\end{enumerate}
For each simulation, around $1\times 10^9$ MC steps were attempted.

\subsection{Bulk canonical MD simulations}

MD simulations of the short-ranged system were performed with the LAMMPS simulation package \cite{Thompson2022}. 
Source code to implement the pair potential with short-ranged
electrostatics is openly available  \cite{github3,Cox2020}.
Cubic simulation boxes with periodic boundary conditions were used. The bulk systems contained 2000 ion pairs, with the length of the box adjusted to yield the desired bulk densities. Dynamics were
propagated using the velocity Verlet algorithm with a time-step of 1 fs. The temperature was maintained using a Langevin thermostat \cite{Schneider1978}. The system was equilibrated for 50 ps and production
runs were performed for at least 2 ns in the NVT ensemble.

To describe the hard-core part of the pair interaction
in MD simulations, as done in Refs.~\onlinecite{Hartel2015, Hartel2016}, we replace the hard-sphere potential with the continuous pseudo hard-sphere potential
\begin{equation}
    u_{\mrm{pHS}}(r) = \begin{cases}
      50 \left(\dfrac{50}{49}\right)^{49} \varepsilon \left[ \left(\dfrac{\sigma}{r}\right)^{50} - \left(\dfrac{\sigma}{r}\right)^{49} \right] + \varepsilon & \quad   r\leq \left(\frac{50}{49}\right)\sigma\\
      0  & \quad   r>\left(\frac{50}{49}\right)\sigma,
    \end{cases}    
\end{equation}
at $k_{\mrm{B}}T/\varepsilon=1.5$, which was found
to reproduce structural and thermodynamic data
of hard-spheres \cite{Jover2012}. The full short-ranged potential becomes
\begin{equation}
 u_{ij, \mrm{R}}(r) =
    \begin{cases}
      u_{\mrm{pHS}}(r)  + \dfrac{q_i q_j}{\epsilon } \dfrac{\mrm{erfc(\kappa r)}}{r} & \quad    r \leq r_{\mrm{c}} \\
      0 & \quad   r>r_{\mrm{c}}.
    \end{cases}       
\end{equation}

\section{Simulations of the long-ranged system} 

\subsection{Bulk GCMC simulations}

For the full long-ranged system, we performed GCMC simulations with \texttt{DL\_MONTE} \cite{Brukhno2021}. Long-ranged electrostatics were treated with Ewald summation \cite{ deLeeuw1997}.
 As insertion and deletion moves can only be done
 with pairs of ions effectively, we controlled $\mu_{\mrm{pair}} = \mu_+ + \mu_-$. Cubic simulation boxes with
 periodic boundary conditions were used.
 The system was equilibrated for $1\times10^4 $ MC steps 
 and $1\times 10^8 $ MC steps were attempted in the production run.

\subsection{Canonical MD simulations}

MD simulations of the long-ranged system were performed with the LAMMPS simulation package \cite{Thompson2022}. 
Cubic simulation boxes with periodic boundary conditions were used. The full pair potential employed in MD is
\begin{equation}
 u_{ij, \mrm{R}}(r) =
    \begin{cases}
      u_{\mrm{pHS}}(r)  + \dfrac{q_i q_j}{\epsilon } \dfrac{1}{r} & \quad    r \leq r_{\mrm{c}} \\
      0 & \quad   r>r_{\mrm{c}}.
    \end{cases}       
\end{equation}
The bulk systems contained 2000 ion pairs, with the length of the box adjusted to yield the desired bulk densities. Dynamics were
propagated using the velocity Verlet algorithm with a time-step of $1\,$fs. The temperature was maintained using a Langevin thermostat \cite{Schneider1978}. The system was equilibrated for $50\,$ps and production
runs were performed for at least $2\,$ns in the NVT ensemble.

Long-ranged electrostatic interactions were evaluated using particle--particle particle--mesh Ewald
summation \cite{hockney1988} (using a $1\,$nm cut off in real space) such that the RMSE in the forces was a factor of $10^5$
smaller than the force between two unit charges separated by a distance of $0.1\,$nm \cite{Kolafa1992}.

\section{Connecting the short-ranged and long-ranged systems}

\subsection{Structure}

\begin{figure}[H]     \centering
  \includegraphics[width=0.9\linewidth]{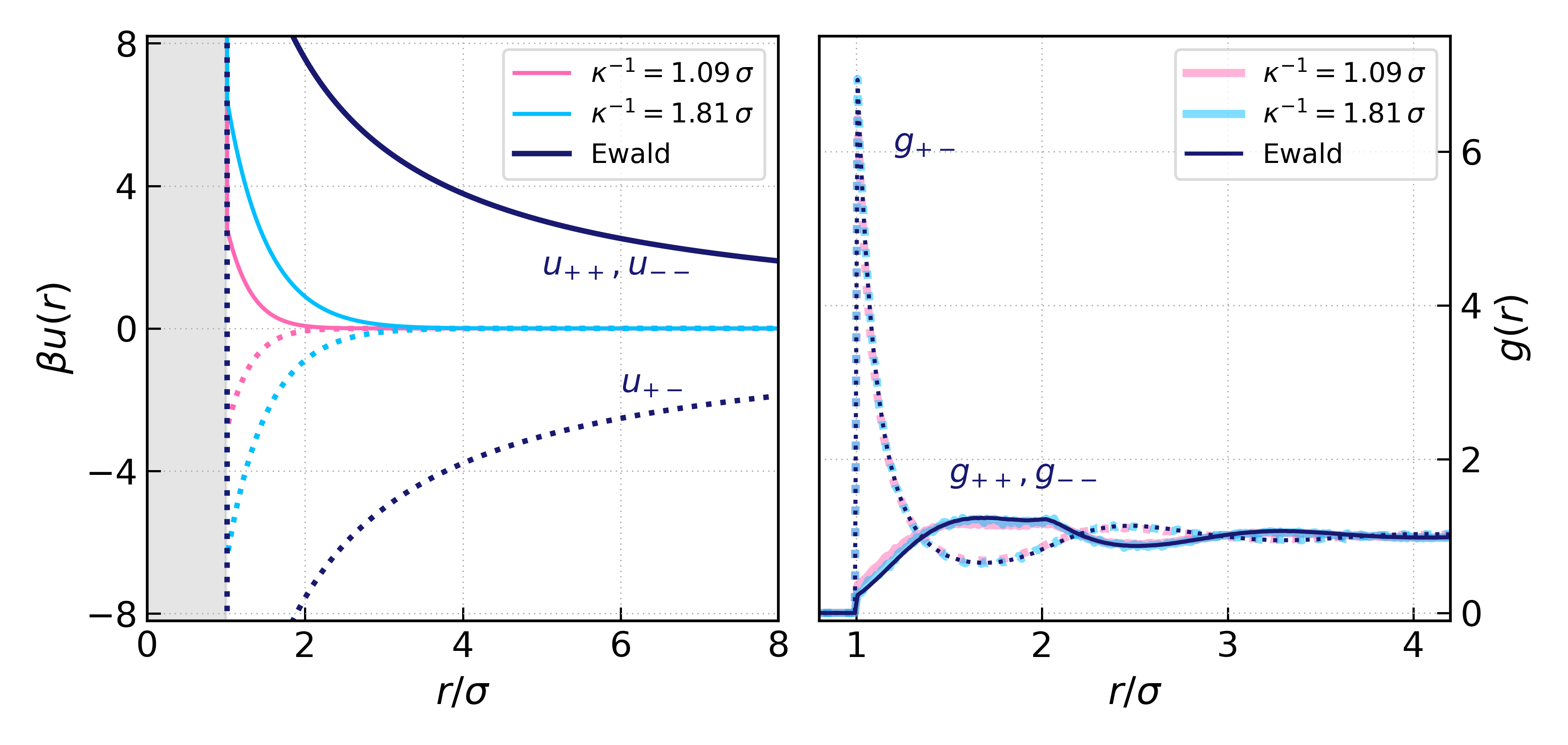}
  \caption{\textbf{Structure of the short-ranged and long-ranged RPM.} (a) The pair potential and (b) radial distribution function of a bulk fluid at the same density $\rho_+=\rho_-=0.315\,\sigma^{-3}$ obtained from GCMC simulations. The cation--cation and anion--anion functions are shown in solid lines while cation--anion functions are in dotted lines.}
  \label{si_structure}
\end{figure}

The pair potentials for the long-ranged and mimic
short-ranged systems are described in
Eq.~\ref{eqn:pair_potential_LR} and
Eq.~\ref{eqn:pair_potential_SR} and
plotted in Fig.~\ref{si_structure}(a).
With $\kappa^{-1}$ chosen to be large enough,
the mimic system can faithfully capture
the short-ranged local structure of the full
system in the bulk, reflected in the 
radial distribution functions in Fig.~\ref{si_structure}(b).
For the bulk structure, discrepancies between the mimic and full system would
only appear at longer wavelength,
as shown in the structure factors in the main paper.

\subsection{Liquid--vapor coexistence}

For the long-ranged RPM, the liquid--vapor
binodal curve has been determined \cite{Yan1999} with the critical temperature
$T^*_{\mrm{c}}=0.0492$. Short-ranged
fluids are expected to have a lower
critical temperature than their long-ranged
counterpart \cite{Johnson1993}.
Indeed, we verified this by constructing
the binodal curve for two short-ranged
mimic systems from GCMC simulations, as shown in Fig.~\ref{si_phase_energy}(a).
In this work, we focus on the isothermal
behavior at $T^*=0.066$, which is supercritical for both the full RPM and the mimic system.

\subsection{Energy}

\begin{figure}[H]     \centering
  \includegraphics[width=0.8\linewidth]{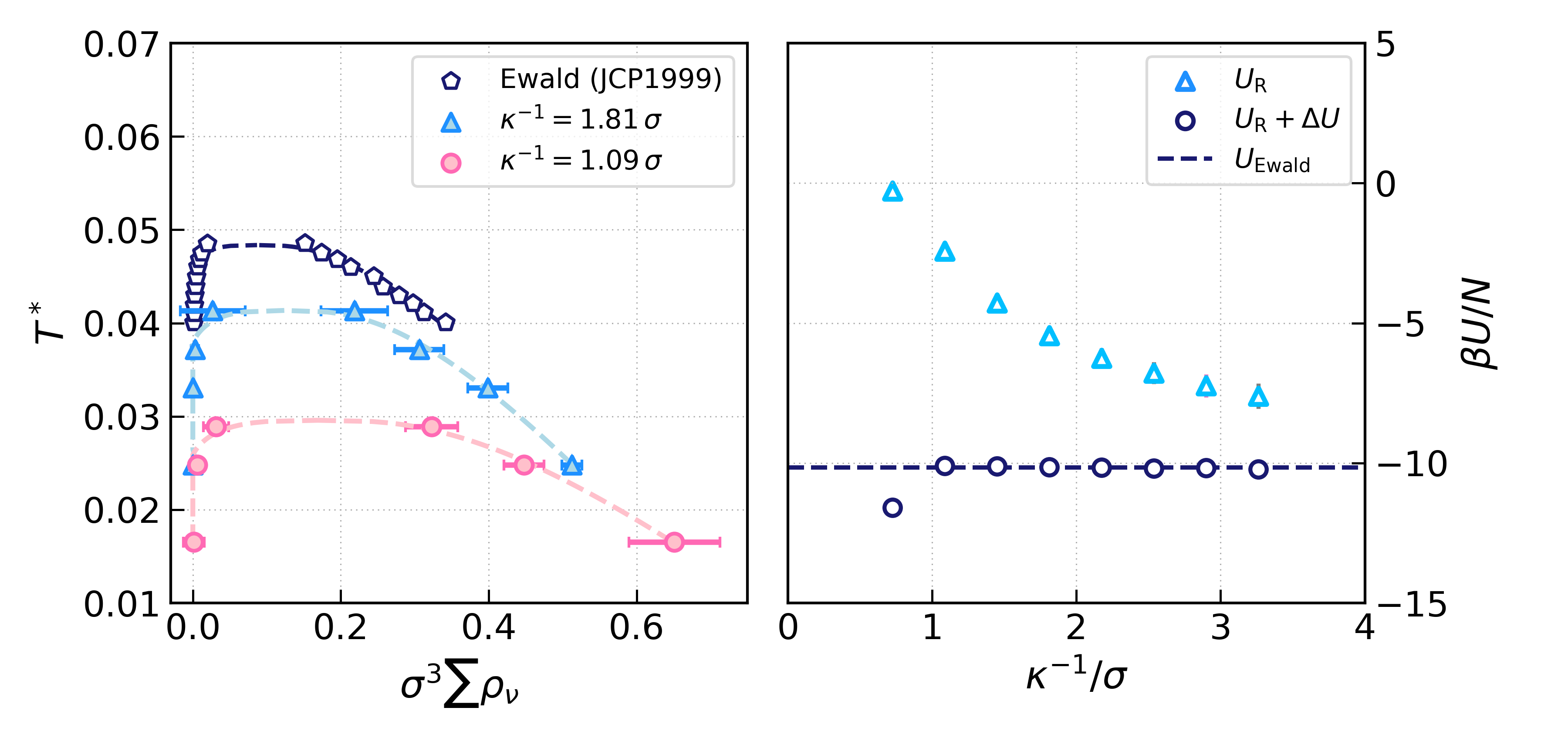}
  \caption{\textbf{Phase diagram and potential energy 
  long-range correction of the RPM.} (a) The liquid--vapor coexistence
  curve of the long-ranged system is from Ref.~\onlinecite{Yan1999}, 
  while those of the short-ranged mimic systems are determined 
  in this work. (b) The total potential energy of the mimic short-ranged
  system at various $\kappa^{-1}$ without (triangles) and with the 
  long-range correction (circles) for the bulk system at $\rho_{\mrm{R},+}=\rho_{\mrm{R},-}=0.263\,\sigma^{-3}$.
  The energy determined from Ewald summation is indicated by the horizontal dashed 
  line.}
  \label{si_phase_energy}
\end{figure}

In Ref.~\onlinecite{Rodgers2009}, an analytical correction
for the Coulombic energy between the long-ranged and short-ranged bulk systems based on the Stillinger--Lovett moment conditions
is derived. For a binary salt, this long-range correction
reduces to
\begin{equation}
     \Delta U \equiv U - U_{\mrm{R}} = - \frac{1}{\kappa^{-1}\sqrt{\pi}}\left(N_{+}q^2_{+} + N_{-}q^2_{-}\right)
     + \frac{1}{\kappa^{-3}\sqrt{\pi}}\frac{V}{2\pi\beta} ,
\end{equation}
where $N_+$ is the number of cations, $N_{-}$ is the number of anions and $V$ is the volume of the system.
As seen in Fig.~\ref{si_phase_energy}, for mimic systems
with $\kappa^{-1} > \sigma$, the total potential
energy with this correction reproduces the energy
of the full system (calculated with Ewald summation)
well.

\subsection{Chemical potential}

The change in internal energy of the bulk system is given as 
\begin{equation}
    \mrm{d}U = T\mrm{d}S + \mu_+\mrm{d}N_{+} + \mu_-\mrm{d}N_{-} - P\mrm{d}V,
\end{equation}
where  $S$ is the entropy and $P$ is the pressure.
The chemical potentials of the anion $\mu_{-}$ and cation
$\mu_{+}$ can be expressed thermodynamically as
\begin{equation}
\begin{split}
        \mu_{-} &= \left( \frac{\partial U}{\partial N_-} \right)_{\mu_{+},V,T} - T\left( \frac{\partial S}{\partial N_-} \right)_{\mu_{+},V,T}, \\
        \mu_{+} &= \left( \frac{\partial U}{\partial N_+} \right)_{\mu_{-},V,T} - T\left( \frac{\partial S}{\partial N_+} \right)_{\mu_{-},V,T}.
\end{split}
\end{equation}
Since the mimic system is constructed to capture the local order and structural variations 
that are expected to dominate the entropy, we approximate
 $S\approx S_{\mrm{R}}$, giving:
 \begin{equation}
     \begin{split}
         \Delta \mu_{-} & \equiv \mu_{-} - \mu_{\mrm{R}, -} = \left( \frac{\partial \Delta U}{\partial N_-} \right)_{\mu_{+},V,T} =  - \frac{q^2_{-}}{\kappa^{-1}\sqrt{\pi}}, \\
         \Delta \mu_{+} & \equiv \mu_{+} - \mu_{\mrm{R}, +} = \left( \frac{\partial \Delta U}{\partial N_+} \right)_{\mu_{-},V,T} =  - \frac{q^2_{+}}{\kappa^{-1}\sqrt{\pi}}. 
     \end{split}
 \end{equation}
For a range of $\kappa^{-1}$,
we find the chemical potentials that gives a
bulk fluid of density $\rho_{\mrm{R},+}=\rho_{\mrm{R},-}=0.263\,\sigma^{-3}$
by running GCMC simulations with various $\mu_{\mrm{R},-}=\mu_{\mrm{R},+}$. As shown in
Fig.~\ref{si_mu_pressure}(a), for $\kappa^{-1}\gtrsim\sigma$, the corrected chemical potentials of the mimic 
system agree well with that of the full system. The long-ranged GCMC simulations of the RPM
were done
with insertions and deletions of cation-ion pairs so
the chemical potential is 
$\mu_{-}=\mu_{+}=\mu_{\mrm{pair}}/2$.

\subsection{Pressure}

\begin{figure}[H]     \centering
  \includegraphics[width=0.8\linewidth]{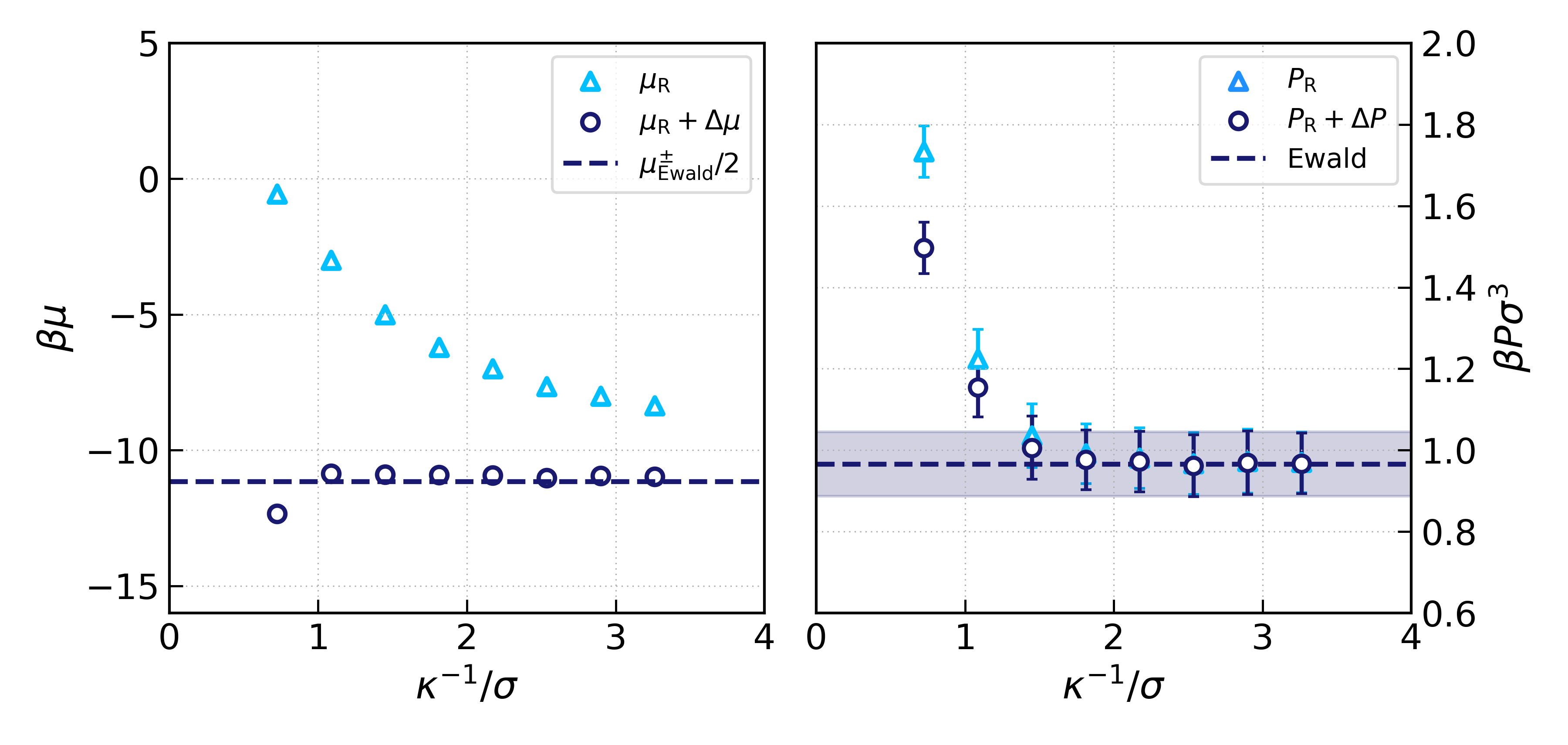}
  \caption{\textbf{Long-range corrections for the chemical potential and pressure for the RPM.} 
  (a) The chemical potential for an ion  and (b)  the pressure of the mimic
  short-ranged system at 
  at various $\kappa^{-1}$ without (triangles) and with the 
  long-range correction (circles) for the bulk system at $\rho_{\mrm{R},+}=\rho_{\mrm{R},-}=0.263\,\sigma^{-3}$.
  The results for the full system (determined with Ewald sums) are indicated by the horizontal dashed lines, with uncertainty in the pressure indicated by the shaded region.
  }
  \label{si_mu_pressure}
\end{figure}

The pressure can be expressed thermodynamically as
\begin{equation}
    P = T \left( \frac{\partial S}{\partial V} \right)_{N_{+},N_{-},T} - \left( \frac{\partial U}{\partial V} \right)_{N_{+},N_{-},T}.
\end{equation}
Again,  $S\approx S_{\mrm{R}}$, leaving the correction for the bulk pressure, consistent with Ref.~\onlinecite{Rodgers2009}, as
\begin{equation}
    \Delta P \equiv P - P_{\mrm{R}} = - \left( \frac{\partial \Delta U}{\partial V} \right)_{N_{+},N_{-},T}  = - \frac{1}{2 \pi^{3/2}\kappa^{-3}\beta}.
\end{equation}
The pressure obtained from simulations of the mimic systems
is shown in Fig.~\ref{si_mu_pressure}(b).
The analytical pressure correction brings the 
result closer to the pressure of the long-ranged 
system. For dense systems,
its contribution is relatively small.

\section{Finite field simulations} \label{sec:finitefield} 

The Hamiltonian of the system under an imposed uniform $\mbf{E}$ field is given as
\begin{equation}
    \mcl{H}_{\mbf{E}}(\mbf{r}^N,\mbf{p}^N) = \mcl{H}_{\mrm{PBC}}(\mbf{r}^N,\mbf{p}^N) - V \mbf{E}\cdot\mbf{P}(\mbf{r}^N),
\end{equation}
where $\mcl{H}_{\mrm{PBC}}$ denotes the Hamiltonian of the periodic
system computed using standard Ewald summation techniques
and $\mbf{P}$ denotes the total polarization formally
defined as the time integral of the current 
\begin{equation}
    \mbf{P}(\mbf{r}^N) = \frac{1}{V}\sum_{i} q_{i} \mbf{r}_i.
\end{equation}
In our simulations, we use an ``electrolyte centered cell'' such that the above 
amounts to a straightforward calculation of the total dipole moment 
of the simulation cell\cite{Zhang2016b,sayer2017,sayer2019}.
In MD simulations, this amounts to adding a force $\mbf{f}_{i}=q_i\mbf{E}$ to each ion $i$ in the simulation.

The finite field Hamiltonian for the system under
constant displacement field $\mbf{D}$ is 
\begin{equation}
    \mcl{H}_{\mbf{D}} (\mbf{r}^N,\mbf{p}^N) =    \mcl{H}_{\mrm{PBC}}(\mbf{r}^N,\mbf{p}^N) + \frac{V}{8\pi}[\mbf{D} - 4\pi\mbf{P}(\mbf{r}^N)]^2.
\end{equation}
The displacement field is related to the electric field through
\begin{equation}
    \mbf{D} = \mbf{E} + 4\pi\mbf{P}.
\end{equation}
In MD simulations, this amounts to adding a field-dependent force to each ion $\mbf{f}_{i}=q_i(\mbf{D}-4\pi\mbf{P})$.

\begin{figure}[H]     \centering
  \includegraphics[width=0.8\linewidth]{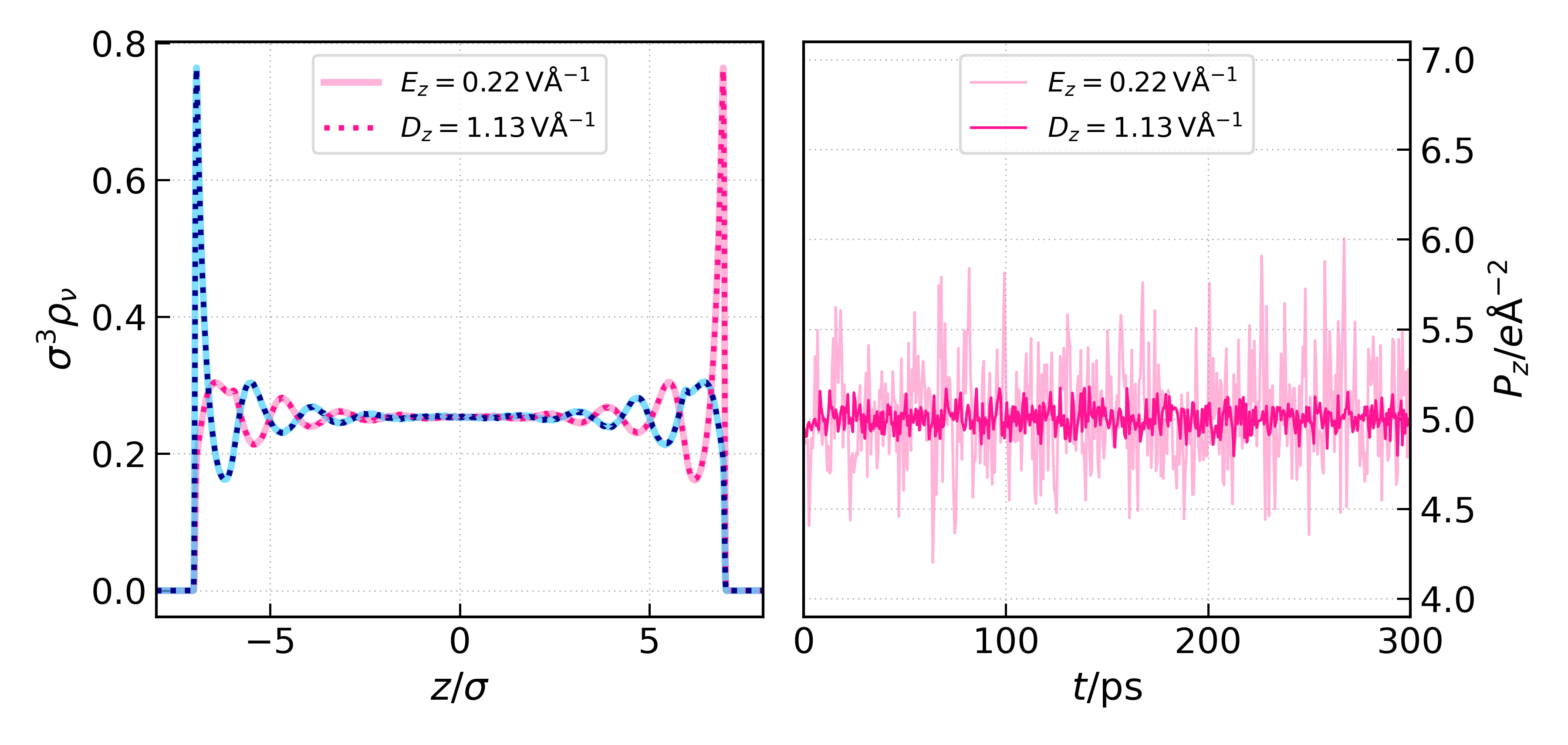}
  \caption{\textbf{Finite field simulations of the RPM confined between two oppositely charged walls.} 
  (a) Ion density profiles obtained are identical
  under a constant electric field 
  $E_z=0.22\,\mrm{V\,\AA^{-1}}$ or under
  a constant displacement field $D_z=1.13\,\mrm{V\,\AA^{-1}}$. (b) The polarization
  $P_z$ in the two simulations are on average the same,
  with smaller fluctuations under constant $D_z$.} 
\end{figure}

\section{Distribution functions} \label{sec:distribution_functions}

We denote the 3D spatial Fourier transform of function $f(\mbf{r})$ as
\begin{equation}
\hat{f}(\mbf{k}) = \int\!\!\mrm{d}\mbf{r}\,f(\mbf{r})\exp(-i\mbf{k}\cdot\mbf{r}),
\end{equation}
and the inverse Fourier transform of $\hat{f}(\mbf{k})$ as
\begin{equation}
f(\mbf{r}) = \frac{1}{(2\pi)^3}\int\!\!\mrm{d}\mbf{k}\,\hat{f}(\mbf{k})\exp(i\mbf{k}\cdot\mbf{r}).
\end{equation}
For a radially symmetric function $f(r)$, the corresponding expressions become
\begin{equation}
 \hat{f}(k) = \frac{4\pi}{k}\int^\infty_0\!\!\mrm{d}r\,rf(r) \sin(kr),
 \label{eqn:radialFT}
\end{equation}
and
\begin{equation}
f(r) = \frac{1}{2\pi^2 r} \int^\infty_0\!\!\mrm{d}k\,k \hat{f}(k)\sin(kr).
 \label{eqn:radialinvFT}
\end{equation}
The planar projection of a radially symmetric
function $f(r)$ is
\begin{equation}
    \overline{f}(z) = \int\!\mrm{d}x\int\!\mrm{d}y\,f(r = \sqrt{x^2 + y^2 + z^2}) = 2\pi\!\int_z^\infty\!\mrm{d}r\, rf(r),
    \label{eqn:planarprojection}
\end{equation}
and the inverse transformation is
\begin{equation}
    f(r) = -\left.\frac{1}{2\pi r} \frac{\partial \overline{f}(z)}{\partial z}\right\vert_{z=r}.
\end{equation}

For a bulk mixture containing $N_{\nu}$ of species $\nu$, we extracted the radial distribution functions  
\begin{equation}
    g_{\nu\lambda}(r) = \frac{\rho_{\mrm{tot}}}{\rho_\nu \rho_\lambda} \left\langle \frac{1}{N} \sum^{N_\nu}_i  \sum^{N_\lambda}_j \delta(r-|\mbf{r}_i-\mbf{r}_j|) \right\rangle,
    \label{eqn:gr}
\end{equation}
where $\rho_{\mrm{tot}}=\sum_{\nu}\rho_{\nu}=\sum_{\nu}N_{\nu}/V$, by generating a histogram of intermolecular distances and appropriate normalization. The corresponding total correlation functions are therefore
\begin{equation}
    h_{\nu\lambda}(r) = g_{\nu\lambda}(r) - 1.
    \label{eqn:hr}
\end{equation}
The high-wavevector part of the partial structure factors
are computed from
\begin{equation}
    S_{\nu\lambda}(k) 
    = \frac{\rho_\nu}{\rho_{\mrm{tot}}}\delta_{\nu\lambda} + \frac{\rho_\nu \rho_\lambda}{\rho_{\mrm{tot}}} \hat{h}_{\nu\lambda}(k).
    \label{eqn:Sk-hk}
\end{equation}
In the low-wavevector limit, we also calculate the structure factor directly from the simulation
trajectories using \cite{Sedlmeier2011}
\begin{equation}
\begin{split}
    S_{\nu\lambda}(k) & =  \left\langle \frac{1}{N}\sum^{N_\nu}_i \sum^{N_\lambda}_j \exp[-i\mbf{k}\cdot(\mbf{r}_{i}-\mbf{r}_{j})]\right\rangle \\
    & = \frac{1}{N}  \left\langle \left[ \sum^{N_{\nu}}_{i}\sin(\mbf{k}\cdot\mbf{r}_i)\right]\left[ \sum^{N_{\lambda}}_{i}\sin(\mbf{k}\cdot\mbf{r}_i)\right] + \left[ \sum^{N_{\nu}}_{i}\cos(\mbf{k}\cdot\mbf{r}_i)\right]\left[ \sum^{N_{\lambda}}_{i}\cos(\mbf{k}\cdot\mbf{r}_i)\right]\right\rangle.
\end{split}
\label{eqn:Sk}
\end{equation}

The Ornstein--Zernike equation that relates the two-body direct correlation functions to the total correlation function for a homogeneous mixture is 
\begin{equation}
    h_{\nu\lambda}(r) = c^{(2)}_{\nu\lambda}(r) + \sum^{n}_{\gamma} \rho_{\gamma} \!\int\!\!\mrm{d}\mbf{r}^\prime c^{(2)}_{\nu\gamma}(|\mbf{r}-\mbf{r}^\prime|) \,h_{\gamma\lambda}(|\mbf{r}^\prime|).
\end{equation}
It is convenient to define two matrices with components \cite{Baxter2003}
\begin{equation}
    \hat{H}_{\nu\lambda}(k) = (\rho_\nu \rho_\lambda)^{1/2} \hat{h}_{\nu\lambda}(k),
    \label{eqn:matrix-H}
\end{equation}
\begin{equation}
    \hat{C}_{\nu\lambda}(k) = (\rho_\nu \rho_\lambda)^{1/2} \hat{c}^{(2)}_{\nu\lambda}(k),
    \label{eqn:matrix-C}
\end{equation}
allowing the Ornstein--Zernike equation to be recast in a matrix form
\begin{equation}
    \hat{\mbf{H}}(k) = \hat{\mbf{C}}(k) +  \hat{\mbf{C}}(k)\hat{\mbf{H}}(k).
\end{equation}
Upon rearrangement, 
\begin{equation}
 \hat{\mbf{C}}(k) = \mbf{I} - [\mbf{I}+\hat{\mbf{H}}(k)]^{-1},
   \label{eqn:matrix-solve}
\end{equation}
where $\mbf{I}$ is the identity matrix.

\begin{figure}[H]     \centering
  \includegraphics[width=\linewidth]{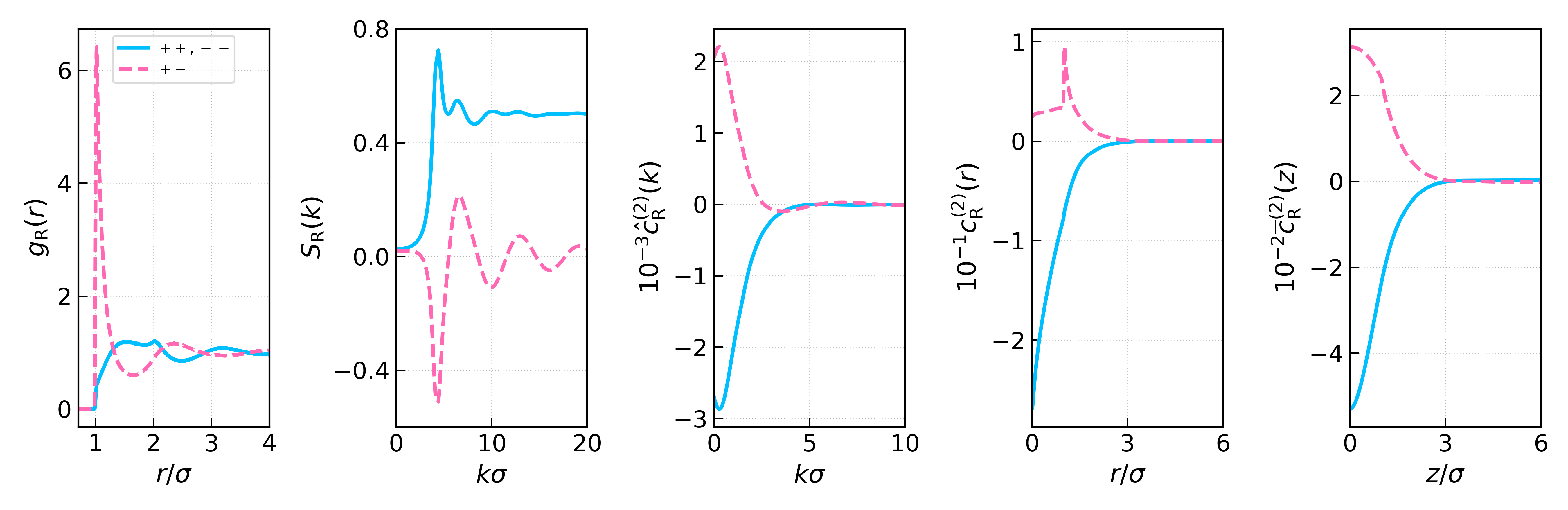}
  \caption{\textbf{Distribution functions.} For a short-ranged mimic bulk system at $\rho_{\mrm{R},+}=\rho_{\mrm{R},-}=0.315\,\sigma^{-3}$, we show the
  (a) radial distribution functions,
  (b) partial structure factors, 
  (c) two-body direct correlation functions
  in reciprocal space, (d) in real space and 
  (e) their planar projections.} 
  \label{si_distribution_functions}
\end{figure}

To obtain the planar bulk two-body direct correlation functions
$\overline{c}^{(2)}_{\mrm{R}, \nu\lambda}(z; \{\rho_{\mrm{b}}\})$,
where $\rho_{\mrm{b}}=\rho_{R,\pm}$ denotes the bulk ionic density, from simulations for bulk
correlation matching, we calculate in order:
\begin{itemize}
    \item $S_{\mrm{R},\nu\lambda}(k; \{\rho_{\mrm{b}}\})$ at the low-wavevector end
    ($k\sigma<5$) directly via Eq.~\ref{eqn:Sk},
    \item $g_{\mrm{R},\nu\lambda}(r; \{\rho_{\mrm{b}}\})$ by generating histograms of 
    interatomic distances via Eq.~\ref{eqn:gr},
    \item $h_{\mrm{R},\nu\lambda}(r; \{\rho_{\mrm{b}}\})$ via Eq.~\ref{eqn:hr},
    \item $\hat{h}_{\mrm{R},\nu\lambda}(k; \{\rho_{\mrm{b}}\})$ 
    at the low-wavevector end through 
    $S_{\mrm{R},\nu\lambda}(k; \{\rho_{\mrm{b}}\})$ via Eq.~\ref{eqn:Sk-hk}
    and at the high-wavevector end via the radial Fourier transform, Eq.~\ref{eqn:radialFT}, of $h_{\mrm{R},\nu\lambda}(r; \{\rho_{\mrm{b}}\})$,
    \item $\hat{H}_{\mrm{R},\nu\lambda}(k; \{\rho_{\mrm{b}}\})$ via
    Eq.~\ref{eqn:matrix-H},
    \item $\hat{C}_{\mrm{R},\nu\lambda}(k; \{\rho_{\mrm{b}}\})$ by 
    solving the matrix equation Eq.~\ref{eqn:matrix-solve},
    \item $\hat{c}^{(2)}_{\mrm{R},\nu\lambda}(k; \{\rho_{\mrm{b}}\})$ via Eq.~\ref{eqn:matrix-C},
    \item $c^{(2)}_{\mrm{R},\nu\lambda}(r; \{\rho_{\mrm{b}}\})$ 
    via the inverse radial
    Fourier transform, Eq.~\ref{eqn:radialinvFT}, of $\hat{c}^{(2)}_{\mrm{R},\nu\lambda}(k; \{\rho_{\mrm{b}}\})$,
    \item $\overline{c}^{(2)}_{\mrm{R},\nu\lambda}(z; \{\rho_{\mrm{b}}\})$ via
    the planar projection, Eq.~\ref{eqn:planarprojection}, of  $c^{(2)}_{\mrm{R},\nu\lambda}(r; \{\rho_{\mrm{b}}\})$.
\end{itemize}
Some of these functions are shown representatively in Fig.~\ref{si_distribution_functions}.

\section{Training one-body direct correlation neural functionals}

\subsection{Training data}

Following Ref.~\onlinecite{Sammuller2023},
we employ randomized simulation conditions 
by generating external potentials of the form
 \begin{equation}
     V_{\nu}(z) = \sum^4_{n=1} A_n \sin\left( \frac{2\pi n z}{L_{z}} + \theta_n \right) + \sum_n B^{\mrm{lin}}_n(z),
 \end{equation}
 where $A_n$ and $\theta_n$ are randomly selected Fourier coefficients and phases, respectively, and $L_z$ is the simulation box length in the $z$ direction.  
 The phases $\theta_n$  are chosen uniformly in the
 interval [0, $2\pi$), and values of $A_n$ are drawn from a normal distribution with zero mean and a variance of $0.06\,(\mrm{k}_{\mrm{B}}T)^2$. 
 Cubic simulation boxes with $L_z=24\,\mrm{\AA}$ with periodic boundary conditions were used. The second summation denotes up to four piecewise linear functions
 \begin{equation}
     B^{\mrm{lin}}_n(z) =  \begin{cases}
     V_1 + \dfrac{V_2 - V_1}{z_2 - z_1}(z - z_1) & \quad   z_1 < z < z_2\\
      0  & \quad  \text{otherwise},
    \end{cases}    
 \end{equation}
with $0 < z_1 < z_2 < L_z$.
The locations $z_1$ and $z_2$ are distributed uniformly
while $V_1$ and $V_2$ follow from an unbiased 
normal distribution with variance of $\,(\mrm{k}_{\mrm{B}}T)^2$.
Similarly, the external electrostatic potential used
is also randomized with an analogous form
 \begin{equation}
     \phi(z) = \sum^4_{n=1} A^{q}_n \sin\left( \frac{2\pi n z}{L_{z}} + \theta^q_n \right) + \sum_n B^{\mrm{lin},q}_n(z).
 \end{equation}
For a subset of simulations, we explicitly
impose planar hard walls by setting $V_{\nu}(z) = \infty$ for $z < z_{\mrm{w}}$ and $z > L_z - z_{\mrm{w}}$
with $z_{\mrm{w}}=2\,\mrm{\AA}$.
The chemical potential is randomly chosen in the range $\beta\mu_{R,\nu} \in [-10.5, -4.0]$. The thermal wavelengths were set to $\Lambda_\nu=1\,\mrm{\AA}$.

In total, 2400 GCMC simulations were used including cases with 
only non-Coulombic external potentials such that $\phi_\mrm{R}(z)=0$, cases with
only electrostatic external potentials such that $V_{\mrm{R},+}(z)=V_{\mrm{R}, -}(z)=0$,
symmetric cases where $V_{\mrm{R}, +}(z)=V_{\mrm{R},-}(z)$ and $\mu_{\mrm{R},+}=\mu_{\mrm{R},-}$ and asymmetric cases.

For each simulation, the ion density profiles $\rho_{\mrm{R},+}(z)$ and 
$\rho_{\mrm{R},-}(z)$ were
sampled with a grid-spacing $\Delta z=0.03\,\mrm{\AA}$.
The one-body direct correlation profiles are calculated
from
\begin{equation}
\begin{split}
    c^{(1)}_{\mrm{R},-}(z; [\{\rho_{\nu}\}]) &= \ln \left( \Lambda_{-}^3\rho_{\mrm{R},-}(z)\right) +  \beta V_{\mrm{R},-}(z) + \beta q_{-} \phi_\mrm{R}(z) - \beta \mu_{\mrm{R},-}, \\
    c^{(1)}_{\mrm{R},+}(z; [\{\rho_{\nu}\}]) &= \ln \left( \Lambda_{+}^3\rho_{\mrm{R},+}(z)\right) +  \beta V_{\mrm{R},+}(z) + \beta q_{+} \phi_\mrm{R}(z) - \beta \mu_{\mrm{R},+}.
\end{split}
\end{equation}

\subsection{Neural network}

\begin{figure}[H]     \centering
  \includegraphics[width=0.85\linewidth]{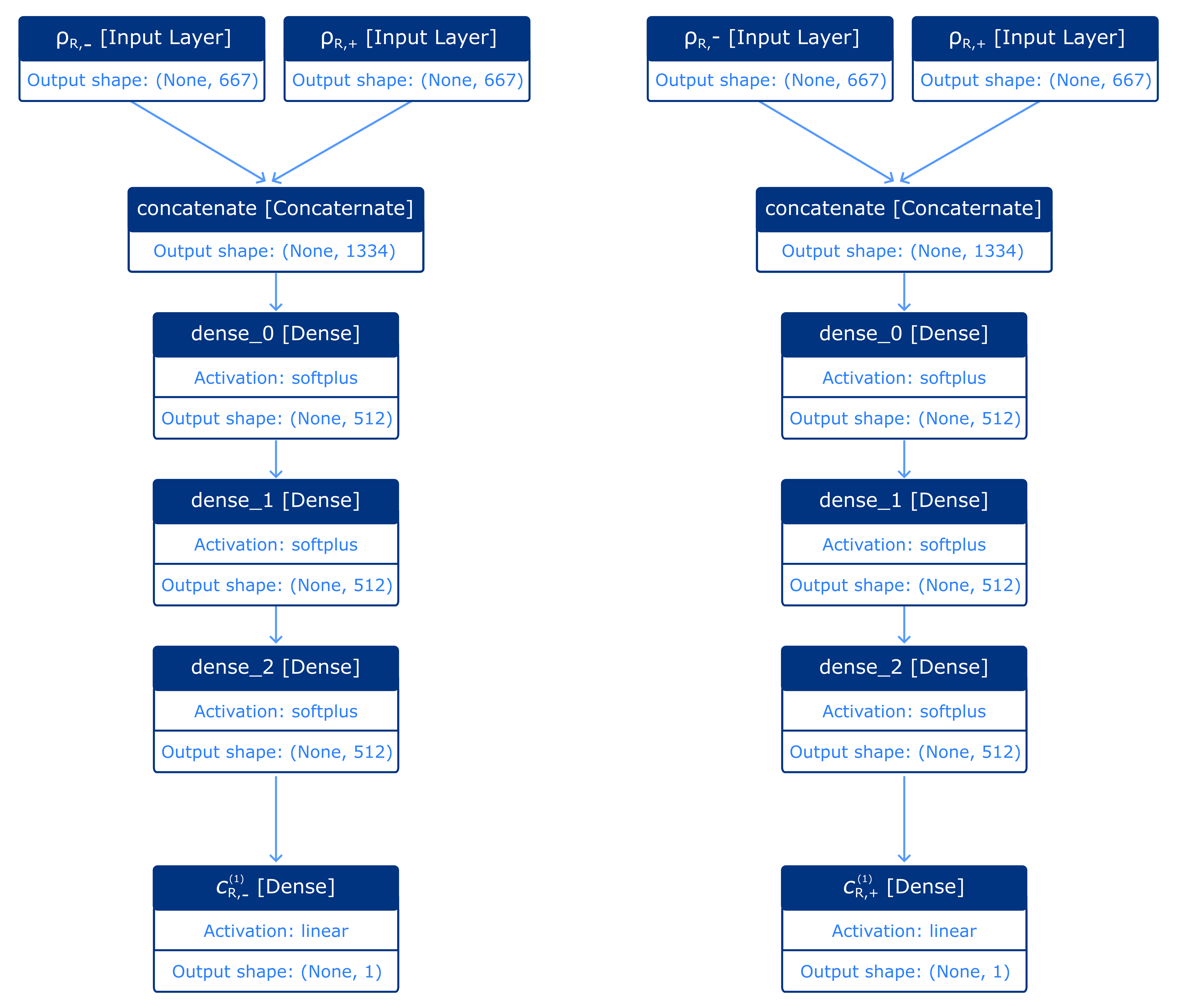}
  \caption{\textbf{Architecture of the neural networks.} 
  Two neural networks are trained to represent 
  the one-body direction correlation functionals
  one for the anion, $c^{(1)}_{\mrm{R},-}(z; [\rho_{\mrm{R},+}, \rho_{\mrm{R},-}])$, and one for the cation, $c^{(1)}_{\mrm{R},+}(z; [\rho_{\mrm{R},+}, \rho_{\mrm{R},-}])$. 
  The output shapes indicate the variable batch size 
  (``None'') and the number of nodes for each layer.} 
  \label{si_ml_model}
\end{figure}

We used two neural networks to represent the one-body
direct correlation functionals, one for $c^{(1)}_{\mrm{R},-}$
and one for $c^{(1)}_{\mrm{R},+}$.
The machine learning routine was implemented in Keras/Tensorflow
with the standard Adam optimizer\cite{Chollet2017}.
The architecture of the models is shown in Fig.~\ref{si_ml_model}.
The input layers take in both ion density profiles,
$\rho_{\mrm{R},+}(z)$ and $\rho_{\mrm{R},-}(z)$, 
in a window of size $10\,\mrm{\AA}$ around the location
of interest. Three fully-connected hidden layers with 512 nodes 
and softplus activation functions then follow,  finally 
resulting in a single scalar value $c^{(1)}_{\mrm{R},-}(z)$
or  $c^{(1)}_{\mrm{R},+}(z)$ at position $z$.
We note that there are also other possible approaches
proposed for learning the short-ranged correlations\cite{Dijkman2024, sammuller2024}.

\subsection{Training procedure}

\begin{figure}[p]     \centering
  \includegraphics[width=0.82\linewidth]{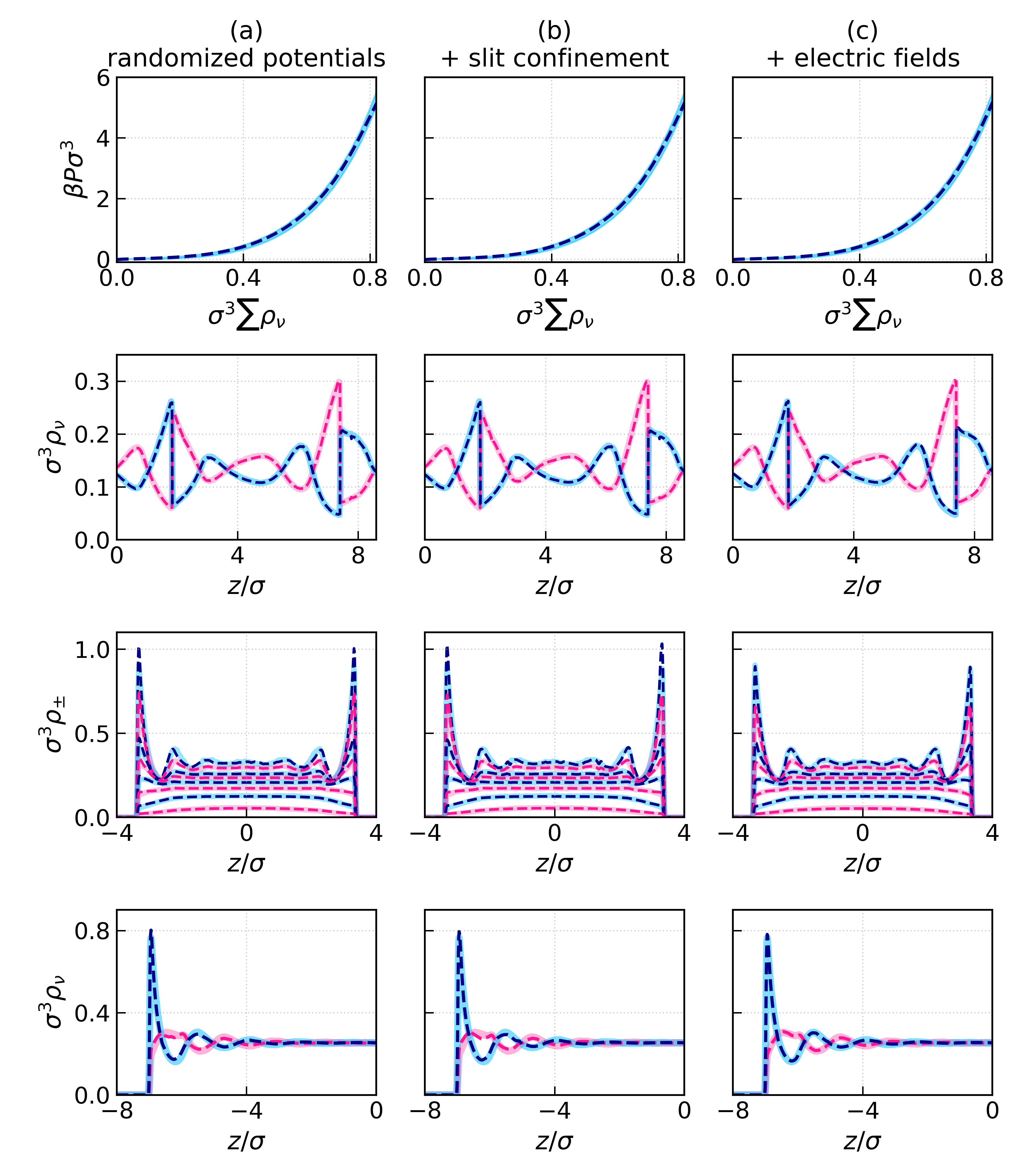}
  \caption{\textbf{Performance of cDFT employing different 
  generations of the neural networks for $\{c^{(1)}_{\mrm{R}}\}$.} Results in dashed lines are shown for neural networks trained with (a)
  randomized potential, (b) with additional slit confinement and (c) with additional electric fields. Overall, good agreement with simulation data (solid lines) is achieved for the equation of state (first row), randomized inhomogeneous profiles (second row), slit confinement (third row) and the electric double layer (final row). The addition of data with electric fields in the final model improves the contact density predictions.} 
  \label{si_compare_models}
\end{figure}

For the first generation of training,
we separate the 2400 independent simulations into 
480 for test set, 480 for validation set and
1440 for training set.
Each model was trained for 200 epochs
in batches of size 256 and
the learning rate was decreased exponentially by 
5\% per epoch from an initial value of 0.001. 
This results in a best mean average error of 0.034
over the validation set for each model,
which is of the same order as the estimated
average noise of the simulation data for 
the one-body direct correlation functions.
Employing this model in cDFT for the full LR system gives the results in Fig.~\ref{si_compare_models}(a).

In addition to these GCMC simulations with randomized potentials, we further add to the data set 128 simulations
in which the SR mimic system experiences the external potential describing the 9-3 Lennard Jones soft walls
at randomized values of the chemical potentials.
With the updated dataset, we then separate the 
independent simulations into 505 for test set, 505 for validation set and 1518 for training set.
Following the same training procedure as before, this results in a best mean average error of 0.038
and employing this model in cDFT for the full LR system gives the results in Fig.~\ref{si_compare_models}(b).

For the last generation of training, we first perform
cDFT for the LR full system under an electric field $E_z$
to generate 72 different data points. From cDFT, we also obtain to the corresponding one-body correlation functions $c^{(1)}_{\mrm{R}, \nu}$
for the mimic system in the process. Using the resulting number density predicted from cDFT, we perform 72 
finite-field MD simulations (see Sec.~\ref{sec:distribution_functions}) of the LR system and collect
the density profiles $\rho_{\nu}=\rho_{\mrm{R},\nu}$.
These 72 combinations of $\{\rho_{\mrm{R},\nu}\}$
and $c^{(1)}_{\mrm{R}, \nu}$ make up the final extra data to the total dataset, which is split into 508 for test set, 508 for validation set and 1524 for training set.
Following the same training procedure as before, the final model results in a best mean average error of 0.032
and employing this model in cDFT for the full LR system gives the results in Fig.~\ref{si_compare_models}(c).

The main computational cost lies in obtaining the training data. The training of the neural networks 
was done on a GPU (NVIDIA GeForce RTX 3060) in a few hours. Once the neural network has been
trained, it can be used relatively inexpensively, with a typical minimization taking 1-2 minutes on a single CPU 
core (and faster on a GPU).

\subsection{Bulk correlation matching}

To reduce noise in the bulk structure predictions, we also
obtained models regularized with bulk two-body direct correlation
functions as proposed in Ref.~\onlinecite{sammuller2024}.
With the bulk simulation of the short-ranged system, we obtain 
the planar bulk two-body direct correlation functions as 
described in Section.~\ref{sec:distribution_functions}. From the neural network representation of the one-body direct correlation functions, 
these are the functional derivatives obtainable via automatic
differentiation
\begin{equation}
\begin{split}
    \overline{c}^{(2)}_{\mrm{R},--}(z; \{\rho_{\mrm{b}}\}) &= \left.\frac{\delta c^{(1)}_{\mrm{R},-}(0;[\{\rho\}])}{\delta \rho_{-}(z)}\right\vert_{\{\rho\}=\{\rho_{\mrm{b}}\}}, \\
    \overline{c}^{(2)}_{\mrm{R},++}(z; \{\rho_{\mrm{b}}\}) &= \left.\frac{\delta c^{(1)}_{\mrm{R},+}(0;[\{\rho\}])}{\delta \rho_{+}(z)}\right\vert_{\{\rho\}=\{\rho_{\mrm{b}}\}}, \\
    \overline{c}^{(2)}_{\mrm{R},+-}(z; \{\rho_{\mrm{b}}\}) &= \left.\frac{\delta c^{(1)}_{\mrm{R},+}(0;[\{\rho\}])}{\delta \rho_{-}(z)}\right\vert_{\{\rho\}=\{\rho_{\mrm{b}}\}}= \left.\frac{\delta c^{(1)}_{\mrm{R},-}(0;[\{\rho\}])}{\delta \rho_{+}(z)}\right\vert_{\{\rho\}=\{\rho_{\mrm{b}}\}}.
\end{split}
\end{equation}

To implement bulk correlation matching, we consider the loss functions
\begin{equation}
\begin{split}
    \mcl{L}_{-} &= \alpha_{\mrm{inhom}}\mcl{L}^{\mrm{inhom}}_{-} + \alpha_{\mrm{pc}}\mcl{L}^{\mrm{pc}}_{-}, \\
    \mcl{L}_{+} &= \alpha_{\mrm{inhom}}\mcl{L}^{\mrm{inhom}}_{+} + \alpha_{\mrm{pc}}\mcl{L}^{\mrm{pc}}_{+},
\end{split}
\end{equation}
in which the losses from local inhomogeneous one-body learning are given as
\begin{equation}
\begin{split}
    \mcl{L}^{\mrm{inhom}}_{-} &= \sum_{i}\left[c^{(\mrm{1})}_{\mrm{R},-}(z;\{\rho\}_{i}) - c^{(\mrm{1})}_{\mrm{R},-, \mrm{sim}}(z;\{\rho\}_{i})\right]^2, \\
    \mcl{L}^{\mrm{inhom}}_{+} &= \sum_{i}\left[c^{(\mrm{1})}_{\mrm{R},+}(z;\{\rho\}_{i}) - c^{(\mrm{1})}_{\mrm{R},+, \mrm{sim}}(z;\{\rho\}_{i})\right]^2,
\end{split}
\end{equation}
and the losses from pair-correlation matching are
\begin{equation}
\begin{split}
    \mcl{L}^{\mrm{pc}}_{-} &= \sum_{i}\left[\overline{c}^{(\mrm{2})}_{\mrm{R},--}(z;\{\rho_{\mrm{b}}\}_{i}) - \overline{c}^{(\mrm{2})}_{\mrm{R},--, \mrm{sim}}(z;\{\rho_{\mrm{b}}\}_{i})\right]^2 + \left[\overline{c}^{(\mrm{2})}_{\mrm{R},+-}(z;\{\rho_{\mrm{b}}\}_{i}) - \overline{c}^{(\mrm{2})}_{\mrm{R},+-, \mrm{sim}}(z;\{\rho_{\mrm{b}}\}_{i})\right]^2, \\
    \mcl{L}^{\mrm{pc}}_{+} &= \sum_{i}\left[\overline{c}^{(\mrm{2})}_{\mrm{R},++}(z;\{\rho_{\mrm{b}}\}_{i}) - \overline{c}^{(\mrm{2})}_{\mrm{R},--, \mrm{sim}}(z;\{\rho_{\mrm{b}}\}_{i})\right]^2 + \left[\overline{c}^{(\mrm{2})}_{\mrm{R},+-}(z;\{\rho_{\mrm{b}}\}_{i}) - \overline{c}^{(\mrm{2})}_{\mrm{R},+-, \mrm{sim}}(z;\{\rho_{\mrm{b}}\}_{i})\right]^2.
\end{split}
\end{equation}
The constant factors $\alpha_{\mrm{inhom}}$ and $\alpha_{\mrm{pc}}$ control the relative
influence of inhomogeneous one-body matching and
pair-correlation matching by weighting the respective loss terms, for which we employ $\alpha_{\mrm{inhom}}=1$ and $\alpha_{\mrm{pc}}=0.0001$.

\section{Electric double layer} 

\begin{figure}[p]     \centering
  \includegraphics[width=0.9\linewidth]{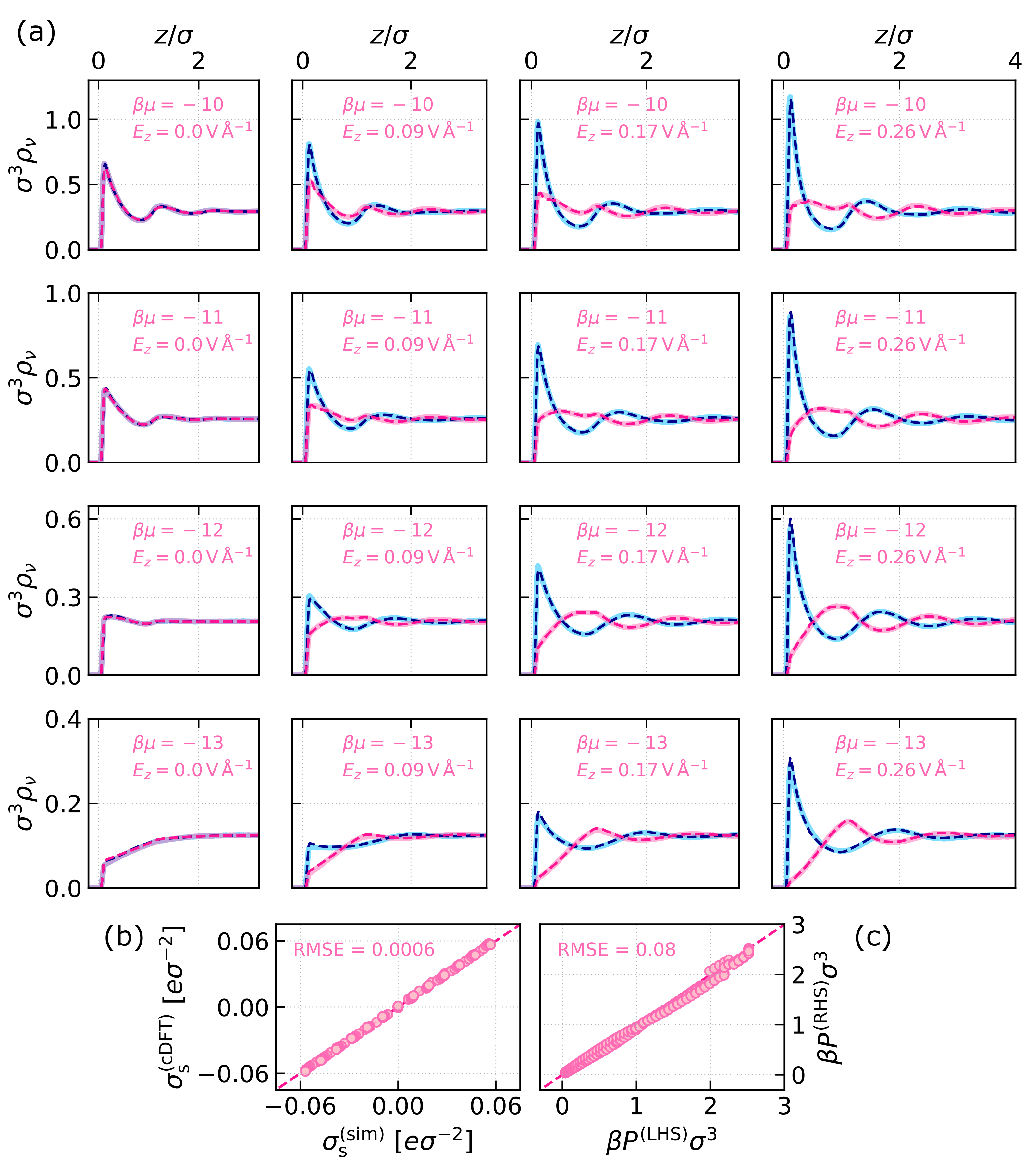}
  \caption{\textbf{Electric double layer of the RPM
  electrolyte in contact with a charged wall.} 
  (a) The equilibrium ion density profiles
  obtained from finite field simulations (solid lines) and 
  from cDFT (dashed lines).
  (b) The surface charge densities on the wall obtained from cDFT agree with that from simulation. 
  (c) The cDFT obeys the contact density theorem in
  Eq.~\ref{eqn:contact_density}, less strictly at
  high density due to the fixed discretization of the
  density profiles obtained from cDFT.} 
  \label{si_contactdensity}
\end{figure}

The formula for the contact density sum rule
for an ionic system at a charged wall
is given as\cite{Henderson1978, Henderson1979, Martin1988}
\begin{equation}
     P = -\sum_{\nu} \int\!\!\mrm{d}z\,\rho_{\nu}(z)\frac{\mrm{d}V_{\nu}(z)}{\mrm{d}z} -  \frac{2\pi \sigma_{\mrm{s}}^2 }{\epsilon},
     \label{eqn:contact_density}
\end{equation}
where $\sigma_{\mrm{s}}$ is the surface charge
density of the wall. In this section, we will examine
the extent to which the cDFT presented in this work obeys Eq.~\ref{eqn:contact_density}.
To this end, we focus on systems with the RPM
confined in a slit made by two walls with
non-electrostatic potential
\begin{equation}
    V_{\nu}(z) = \varepsilon_{\mrm{LJ}} \left[ \frac{2}{15}\left(\frac{\sigma_{\mrm{LJ}}}{z}\right)^9 - \left(\frac{\sigma_{\mrm{LJ}}}{z}\right)^3 \right] +  \varepsilon_{\mrm{LJ}} \left[ \frac{2}{15}\left(\frac{\sigma_{\mrm{LJ}}}{z_{\mrm{min}}}\right)^9 - \left(\frac{\sigma_{\mrm{LJ}}}{z_{\mrm{min}}}\right)^3 \right],
\end{equation}
where ${z_{\mrm{min}}}=\left(\frac{2}{5}\right)^{1/6}\sigma_{\mrm{LJ}}$, $\sigma_{\mrm{LJ}}=\sigma$, $\epsilon_{\mrm{LJ}}=k_{\mrm{B}}T$
under an electric field $E_{z}$.
In the finite field simulations, the two
oppositely charged
walls are not represented explicitly. 
Instead, the surface charge is directly
determined by the displacement field
$D_z$
\begin{equation}
    \sigma^{\mrm{(sim)}}_{\mrm{s}} = \frac{\langle D_z\rangle }{4\pi} = \frac{E_z}{4\pi} +  \langle P_z\rangle.
\end{equation}
This approach assumes that the electrolyte screens perfectly, which is reasonable if the region occupied by the electrolyte is thick enough. 
From the cDFT, we obtain the surface charge by integrating the equilibrium charge densities:
\begin{equation}
    \sigma^{\mrm{(cDFT)}}_{\mrm{s}} = - \int^{L_z/2}_{0}\!\!\mrm{d}z\, \sum_{\nu} q_{\nu}\rho_{\nu}(z).
\end{equation}
We obtained the equilibrium density profiles from cDFT and simulations
for the RPM at $\beta\mu_{-}=\beta\mu_{+} \in [-13.5, -8.5]$ under
$E_z\in[0.0, 0.26]\,\mrm{V\,{\AA}^{-1}}$, a subset of which are shown
in Fig.~\ref{si_contactdensity}(a).  The cDFT performs the best for
density range $\sigma^3\rho_{\pm}\in [0.1, 0.4]$, as expected since
the training data for $c_{\mrm{R}}^{(1)}$ focuses on this density
region. Higher accuracy for lower and higher density would likely be
achieved by extending the training data further.  In
Fig.~\ref{si_contactdensity}(b), we verified that
$\sigma^{\mrm{(cDFT)}}_{\mrm{s}}$ is in agreement with
$\sigma^{\mrm{(sim)}}_{\mrm{s}}$.  Then for each cDFT result, we
evaluate the bulk pressure in Eq.~\ref{eqn:contact_density},
$P^{\mrm{(LHS)}}$, by functional integration together with the
pressure correction as presented in the main paper. The right hand
side of Eq.~\ref{eqn:contact_density}, $P^{\mrm{(RHS)}}$, is evaluated
by numerical integration with the equilibrium ionic profiles from
cDFT, shown in Fig.~\ref{si_contactdensity}(c).

\newpage

\section{Primitive model and multivalent system}

We illustrate the generalizability of our approach to all ionic fluids by considering
the primitive model (PM) and a multivalent system. For the PM, we have changed the sizes of the anion and
cation to $\sigma_- = 4\sigma/3$ and $\sigma_+ = 2\sigma/3$,
respectively. For the multivalent system, we consider equal-sized
anions and cations of diameter $\sigma$, but increase the valency of
the anion such that $|q_-|/|q_+| = 2$. We use the same temperature as for
the RPM, $T^{*}= \epsilon\sigma k_{\mrm{B}}T/|q_{+}q_{-}|=0.066$.

\begin{figure}[H]     \centering
  \includegraphics[width=0.58\linewidth]{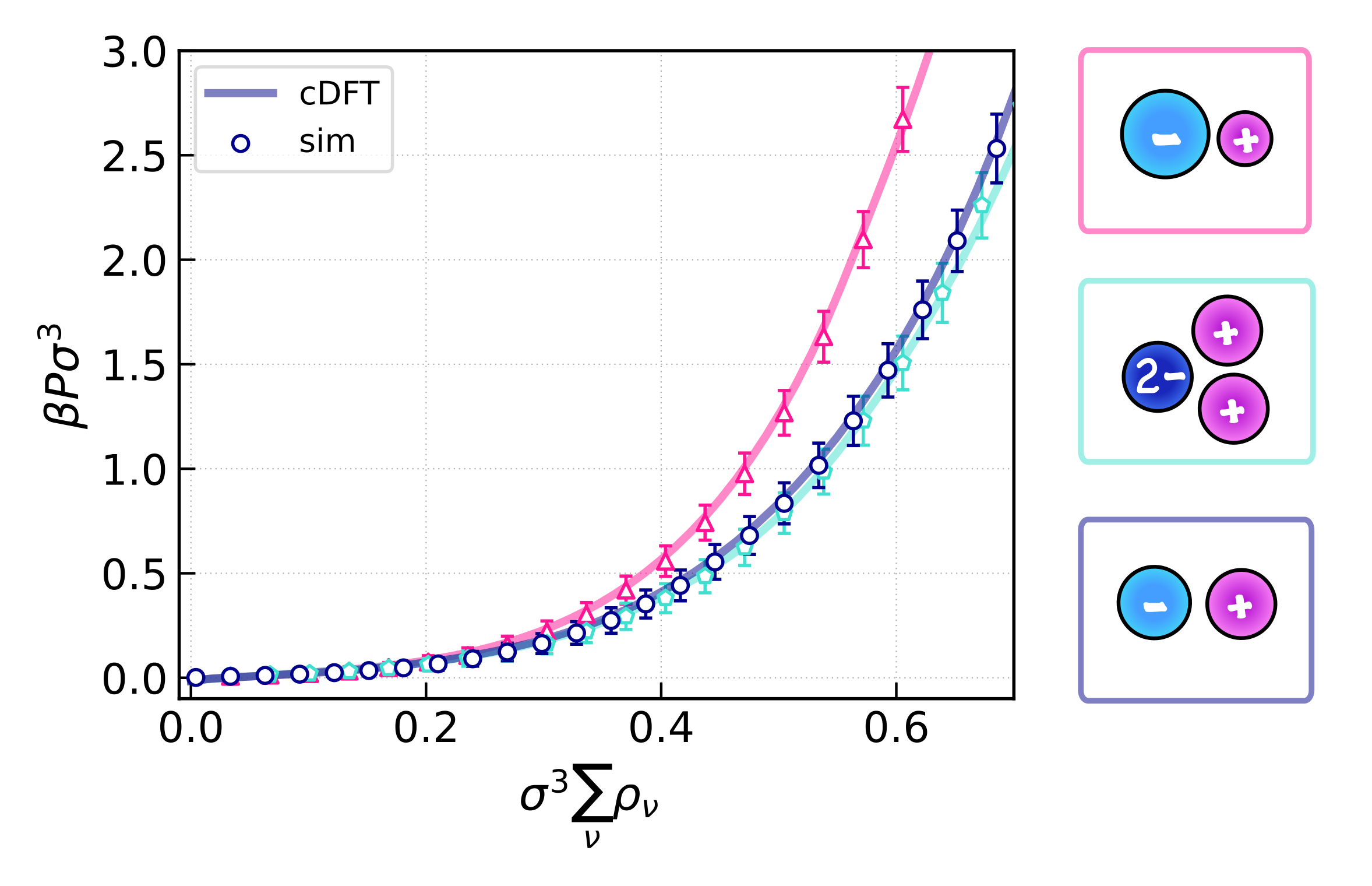}
  \caption{\textbf{The equations of state of additional ionic systems.} Prediction from cDFT is in excellent agreement with simulations both for 
  the restricted primitive model (blue circles), the primitive model (pink triangles) and a  multivalent system (cyan pentagons).}
   \label{si_primitive_models}
\end{figure}

Analogous to the RPM, for each new ionic system, reference data is generated via 
GCMC simulations of $\sim$ 2500 systems with randomized external conditions of
which $\sim$ 1500 are used for training and $\sim$ 500 respectively for validation and testing. 
In addition to inhomogenous profiles of the electric double layers shown in the main paper, 
in Fig.~\ref{si_primitive_models}, we show the results for the equations of state.

\bibliography{references}